\documentclass[
	amsmath,
	amssymb,
	a4paper,
	reprint,	
	longbibliography,
	fleqn,
	showpacs,
	superscriptaddress,
	floatfix
]{revtex4-1}

\usepackage[utf8]{inputenc}
\usepackage[T1]{fontenc}
\usepackage{microtype}

\usepackage{fourier}
\usepackage[mathcal]{euscript}

\usepackage{tabularx}
\usepackage{booktabs}
\usepackage{multirow}

\usepackage{bm}
\usepackage{amssymb}
\usepackage{amsthm}
\usepackage{braket}
\usepackage{mathtools}
\usepackage[bbgreekl]{mathbbol}

\usepackage{graphicx}
\usepackage{dcolumn}
\usepackage[caption=false]{subfig}
\usepackage{color}

\usepackage[version=3]{mhchem}

\usepackage{hyperref}

\definecolor{webgreen}{rgb}{0,.5,0}
\definecolor{webbrown}{rgb}{.6,0,0}
\definecolor{grigio}{rgb}{.85,.85,.85} 
\definecolor{RoyalBlue}{rgb}{0.0, 0.14, 0.4}

\newcommand{\myTitle}{Nonequilibrium thermodynamics of light-induced reactions}

\newcommand{\emanuele}{Emanuele Penocchio}
\newcommand{\riccardo}{Riccardo Rao}
\newcommand{\massimiliano}{Massimiliano Esposito}
\newcommand{\ulAffiliation}{Complex Systems and Statistical Mechanics, Department of Physics and Materials Science, University of Luxembourg, L-1511 Luxembourg City, G.D.~Luxembourg}
\newcommand{\iasAffiliation}{Simons Center for Systems Biology, School of Natural Sciences, Institute for Advanced Study, 08540 Princeton (NJ), U.S.A.}

\newcommand{\dt}{\mathrm{d}_{t}}

\newcommand{\e}{\mathrm{e}} 

\newcommand{\NA}{N_{\mathrm{A}}}

  {\left\lbrace\begin{array}{@{}l@{}}}%
  {\end{array}\right.}

\theoremstyle{definition}

\theoremstyle{definition}

\theoremstyle{definition}

\theoremstyle{plain}

\definecolor{butter1}{rgb}{0.98,0.91,0.31}
\definecolor{butter2}{rgb}{0.93,0.83,0}
\definecolor{butter3}{rgb}{0.77,0.63,0}
\definecolor{skyblue1}{rgb}{0.45,0.62,0.81}
\definecolor{skyblue2}{rgb}{0.2,0.39,0.64}
\definecolor{skyblue3}{rgb}{0.13,0.29,0.53}
\definecolor{scarlet1}{rgb}{0.93,0.16,0.16}
\definecolor{scarlet2}{rgb}{0.8,0,0}
\definecolor{scarlet3}{rgb}{0.64,0,0}
\definecolor{chameleon1}{rgb}{0.54,0.88,0.2}
\definecolor{chameleon2}{rgb}{0.45,0.82,0.09}
\definecolor{chameleon3}{rgb}{0.3,0.6,0.02}
\definecolor{orange1}{rgb}{0.98,0.68,0.24}
\definecolor{orange2}{rgb}{0.96,0.47,0}
\definecolor{orange3}{rgb}{0.8,0.36,0}
\definecolor{plum1}{rgb}{0.68,0.5,0.66}
\definecolor{plum2}{rgb}{0.46,0.31,0.48}
\definecolor{plum3}{rgb}{0.36,0.21,0.4}
\definecolor{chocolate1}{rgb}{0.91,0.72,0.43}
\definecolor{chocolate2}{rgb}{0.75,0.49,0.07}
\definecolor{chocolate3}{rgb}{0.56,0.35,0.01}
\definecolor{aluminium1}{rgb}{0.93,0.93,0.92}
\definecolor{aluminium2}{rgb}{0.82,0.84,0.81}
\definecolor{aluminium3}{rgb}{0.73,0.74,0.71}
\definecolor{aluminium4}{rgb}{0.53,0.54,0.52}
\definecolor{aluminium5}{rgb}{0.33,0.34,0.32}
\definecolor{aluminium6}{rgb}{0.18,0.2,0.21}  

\definecolor{webgreen}{rgb}{0,.5,0}
\definecolor{webbrown}{rgb}{.6,0,0}
\definecolor{grigio}{rgb}{.85,.85,.85} 
\definecolor{RoyalBlue}{rgb}{0.0, 0.14, 0.4}

\hypersetup{%
    colorlinks=true, linktocpage=true, pdfstartpage=1, pdfstartview=FitV,%
    breaklinks=true, pdfpagemode=UseNone, pageanchor=true, pdfpagemode=UseOutlines,%
    plainpages=false, bookmarksnumbered, bookmarksopen=true, bookmarksopenlevel=1,%
    hypertexnames=true, pdfhighlight=/O,
    urlcolor=webbrown, linkcolor=RoyalBlue, citecolor=webgreen, 
    pdftitle={\myTitle},%
    pdfsubject={},%
    pdfkeywords={},%
    pdfcreator={pdfLaTeX},%
    pdfproducer={LaTeX REVTeX}%
}

\begin{document}

\title{\myTitle}

\author{\emanuele}
\email{emanuele.penocchio@uni.lu}
\affiliation{\ulAffiliation}
\author{\riccardo}
\email{riccardorao@ias.edu}
\affiliation{\ulAffiliation}
\affiliation{\iasAffiliation}
\author{\massimiliano}
\email{massimiliano.esposito@uni.lu}
\affiliation{\ulAffiliation}

\begin{abstract}
Current formulations of nonequilibrium thermodynamics of open chemical reaction networks only consider chemostats as free-energy sources sustaining nonequilibrium behaviours.
Here, we extend the theory to include incoherent light as a source of free energy.
We do so by relying on a local equilibrium assumption to derive the chemical potential of photons relative to the system they interact with.
This allows us to identify the thermodynamic potential and the thermodynamic forces driving light-reacting chemical systems out of equilibrium.
We use this framework to treat two paradigmatic photochemical mechanisms describing light-induced unimolecular reactions ---~namely the adiabatic and diabatic mechanisms~--- and highlight the different thermodynamics they lead to. Furthermore, using a thermodynamic coarse-graining procedure, we express our findings in terms of commonly measured experimental quantities such as quantum yields.
\end{abstract}

\maketitle

\section{Introduction}

The importance of physicochemical processes involving light is hard to overstate. They are ubiquitous in nature and constitute the prime mechanism driving our planet out of equilibrium. They power for instance the climate dynamics~\cite{lucarini2020} as well as photosynthesis~\cite{calvin1956,Blankenship2014}.
The former example is mostly due to \textit{photophysical} processes~\cite{Balzani2014}, as molecules absorb high frequency photons coming from the sun and decay back to the original ground state by emitting photons in the form of heat.
Photosynthesis instead is a \textit{photochemical} process~\cite{Balzani2014} since it involves chemical reactions powered by light.
Here, the energy of photons is transduced into chemical free energy as molecules with a high chemical potential are synthesised from low-chemical-potential reactants (e.g. glucose from carbon dioxide and water)~\cite{barber2009}.
The opposite process where light is generated from chemical reactions can also happen, as in bioluminescence~\cite{navizet2011}.
Another crucial photochemical reaction senses light in animal vision: the photoisomerization of the 11-cis retinal chromophore to its all-trans form in rhodopsin \cite{wald1963,cerullo2010}.
Nowadays, photochemical reactions are also commonly used to power synthetic molecular machines and devices~\cite{giuseppone2020}.
Examples include:
the first synthetic molecular motor~\cite{feringa1999};
one of the few examples of a synthetic bimolecular motor~\cite{credi2015};
early experimental demonstrations of the ability of molecular motors to move macroscopic objects~\cite{leigh2005,feringa2006};
the first material performing macroscopic work thanks to molecular motors~\cite{giuseppone2015,giuseppone2017};
and the first experimental design of a molecular Maxwell-Demon~\cite{leigh2007}.
Breakthrough experiments in the field of artificial photosynthesis also showed the possibility of using photochemical reactions to harvest free energy from a light source and transduce it into a gradient of ions across a membrane~\cite{moore1997,moore2002,bhosale2006}.

It comes therefore as no surprise that light-reacting matter has long intrigued scientists.
But it is only during the Nineteenth Century that photophysics and photochemistry started to be systematically investigated by pioneers like von Grotthuss~\cite{vongrotthuss1819}, Draper~\cite{draper1843}, Lemoine~\cite{lemoine1895}, and Ciamician~\cite{ciamician1908}.
In the Twentieth Century, the development of modern physics spurred further investigations.
The concept of photon allowed to formulate the so-called Stark–Einstein law~\cite{Rohatgi-Mukherjee1978,Balzani2014}, thus initiating the efforts to establish a correspondence between the number of reacting molecules and the number of photons absorbed in a photochemical reaction.
Moreover, electronic structure theory~\cite{Szabo1996} laid the basis for both the interpretation of absorption spectra of organic molecules and the mechanistic understanding of chemical processes involving light~\cite{Bowen1946,Fleming2010}.
Nowadays, a theory describing photophysical and photochemical processes is well established~\cite{Boggiopasqua2015,Persico2018} and allows for computational studies~\cite{Robb2000,persico2001} and the accurate interpretation of ultra-fast spectroscopic experiments able to unveil mechanistic details~\cite{cerullo2010}.

Focusing on energetic aspects, equilibrium thermodynamic descriptions of light-matter interactions were first formulated by  Kirchhoff~\cite{kirchhoff1860}, Wien~\cite{wien1896}, Reyleigh~\cite{rayleigh1900} and Plank~\cite{plank1901} for black body radiation and by Einstein~\cite{einstein1905} for the photoelectric effect. 
Nonequilibrium aspects only started to be investigated in the second half of the last Century~\cite{ross1966}, often in the context of photosynthesis~\cite{ross1967calvin,ross1967}.
The concept of a chemical potential quantifying the amount of free energy entering the system upon the absorption of a photon was however first introduced in context of semiconductor solid state physics~\cite{wurfel1982,wurfel2005,Wurfel2016}.
Later on, this chemical potential was recognized as the maximal amount of reversible work that may be done by a photon which is absorbed by a chemical system~\cite{ries1991,Milazzo1997}.
However, a nonequilibrium thermodynamics description of how light can drive photophysical and/or photochemical reactions out of equilibrium is still lacking. 
While progress has been made in recent years in describing the nonequilibrium thermodynamics of chemical reactions (in ideal~\cite{gaspard04,qian05,polettini2014,ge16,rao2016,rao18,penocchio2019} and non-ideal solutions~\cite{avanzini2021ni}, and even with diffusion~\cite{falasco2018tp,avanzini2019cw}), no free-energy sources other than chemical potentials gradients (which can be maintained by chemostatting some species) have been considered.

In this paper, we establish the foundations of a nonequilibrium thermodynamic description of elementary chemical processes coupled to incoherent radiation.
We show that the concept of chemical potential of the photons naturally emerges from the assumption of local equilibrium, which lays at the core of the modern formulations of nonequilibrium thermodynamics of chemical reactions~\cite{prigogine1949,Kondepudi2014,rao2016}.
Einstein's relations between absorption and emission coefficients~\cite{einstein1916} provide a local detailed balance principle connecting thermodynamic quantities to dynamical ones. 
This leads to an explicit expression for the entropy production rate of elementary photochemical processes which allows to unambiguously identify the thermodynamic potentials at work in photochemical systems and the forces driving them out of equilibrium.
The outcome is a thermodynamic description of how light and reactions relax to equilibrium when interacting in a closed box or how radiations sources can drive reactions out-of-equilibrium in an open system scenario.
We also connect our theory to quantities commonly measured experimentally such quantum yields~\cite{Montalti2006} using a thermodynamically consistent coarse-graining~\cite{wachtel2018,avanzini2020cg}.
A key finding is that photochemical processes with the same coarse-grained reaction fluxes may have different dissipation rates depending on the underlying elementary mechanism.

This work lays the basis for quantitative energetic considerations in light-powered chemical systems.
A direct application could be to evaluate the thermodynamic efficiency of light-driven molecular motors, whose performance has been assessed just from a dynamical standpoint~\cite{feringa2009,sabatino2019}.
In the long run, our results pave the way for a unified thermodynamic perspective on free-energy transduction from different sources (\textit{e.g.} light, electricity, chemostats) mediated by chemical reactions.

The content of the paper is organized as follows.
In Sec.~\ref{sec:absEmProc}, we introduce our thermodynamic description on a simple model of elementary photophysical reaction.
In Sec.~\ref{sec:photochemistry}, we apply our theory to two common schemes of photochemical reactions, adiabatic and diabatic mechanisms~\cite{Balzani2014}.
The practical relevance of our theoretical results is discussed in Sec.~\ref{sec:discussion}.
Throughout the paper, we consider homogeneous ideal dilute solutions interacting with monochromatic light.
This allows us to keep the mathematical treatment simple.
Future extensions are outlined in Sec.~\ref{sec:conclusions}.

\section{Elementary Photophysical Processes}
\label{sec:absEmProc}

The aim of this section is to build the nonequilibrium thermodynamic description of the prototypical photophysical process, which is depicted in Figure~\ref{FIG_ab:em_elementary}. 
To do so, we adopt the approach developed in \cite{polettini2014,rao2016,penocchio2019} for purely thermal chemical processes.
In \S~\ref{sec:kinetics} we introduce the kinetics of the model.
We then discuss how thermodynamic state functions can be specified for chemical species and radiation in far-from-equilibrium regimes, \S~\ref{sec:thermopot}.
The connection between dynamics and thermodynamics is made in \S~\ref{sec:ldb} through the condition of local detailed balance.
This condition ensures that a closed system made of light-reacting chemicals relaxes to equilibrium and does so by minimizing a suitable thermodynamic potential, \S~\ref{sec:closedsystem}.
Instead, in the presence of external light sources, nonequilibrium steady states can be maintained by flows of free energy crossing the system, \S~\ref{sec:radiostatted}.

\subsection{Kinetics}
\label{sec:kinetics}

\begin{figure}[tb]
    \includegraphics[width=.50\textwidth]{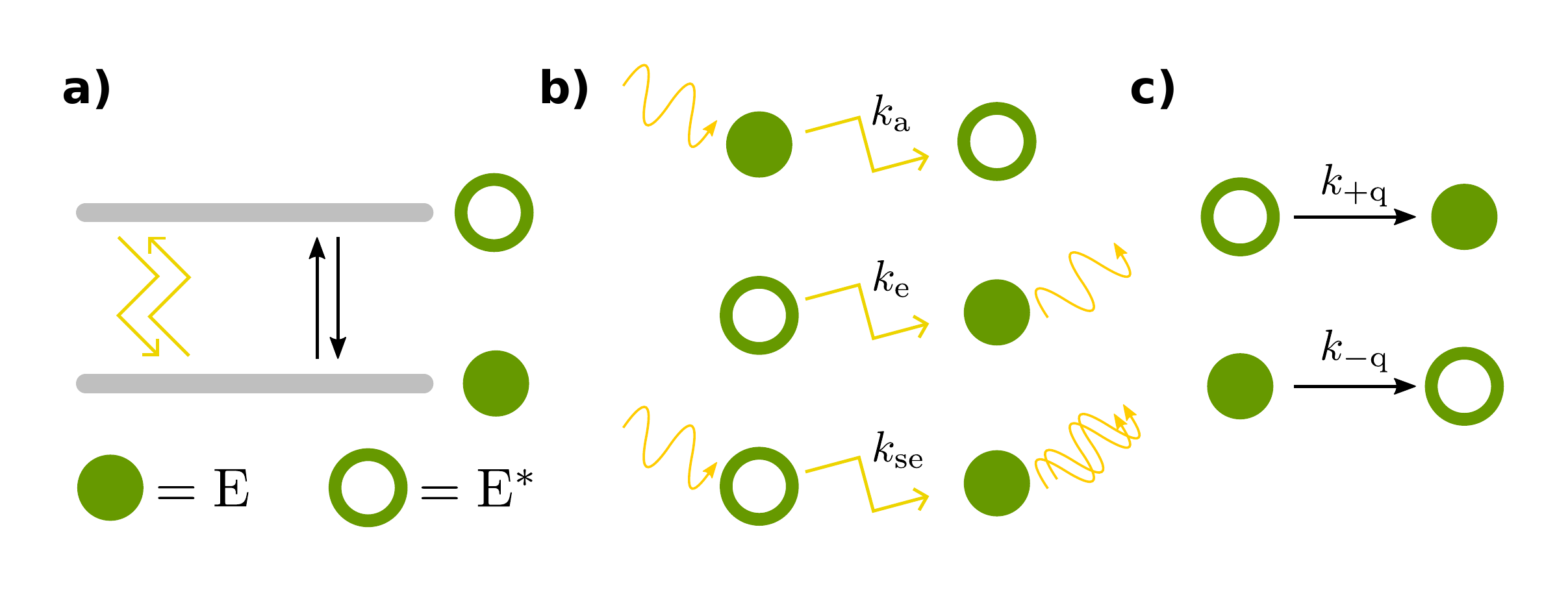}
    \centering
    \caption{\textbf{| Elementary photophysics of a two-level molecule.}
    a) Schematic representation (also known as Jablonski diagram~\cite{Balzani2014}) of a two level molecule undergoing radiative and non-radiative transitions (quenching) among ground ($\ce{E}$) and excited ($\ce{E^\ast}$) electronic states.
    b) Absorption, spontaneous emission and stimulated emission.
    c) Quenching transitions.
}
\label{FIG_ab:em_elementary} 
\end{figure}

We consider the \emph{photophysical mechanism} depicted in Figure~\ref{FIG_ab:em_elementary}.
It comprises the so-called \textit{primary events} in experimental photochemisty~\cite{foerster1973, Balzani2014}, \textit{i.e.} transitions between different electronic states, here denoted as $\mathrm{E}$ and $\mathrm{E^\ast}$.

Among such transitions, those highlighted in Figure~\ref{FIG_ab:em_elementary}b involve light in the form of a photon of frequency $\nu$ ---~denoted $\gamma_{\nu}$.
Their kinetics is characterized by the Einstein coefficients~\cite{einstein1916,Cohen-Tannoudji1998}:
\begin{align}
\mathrm{E} + \gamma_{\nu} & \ce{->[k_{\mathrm{a}}]} \mathrm{E^\ast} & & \textit{absorption}	\label{ce:absorption} \\
\mathrm{E^\ast} & \ce{->[k_{\mathrm{e}}]} \mathrm{E} + \gamma_{\nu} & &
\textit{spontaneous emission} \label{ce:emission} \\
\mathrm{E^\ast} + \gamma_{\nu} &\ce{->[k_{\mathrm{se}}]} \mathrm{E} + 2 \gamma_{\nu} & & \textit{stimulated emission}. \label{ce:stimulatedemission}
\end{align}
Taken together, these three transitions constitute the elementary \emph{photophysical reaction}.
Its net current reads
\begin{equation}
	J_\nu = J_{+\nu} - J_{-\nu} \, ,
	\label{eq:ab--emCurrent}
\end{equation}
where
\begin{subequations}
\begin{align}
    	J_{+\nu} & = k_{\mathrm{a}} n_{\nu} [\mathrm{E}] \\
    	J_{-\nu} & = \left( k_{\mathrm{e}} + k_{\mathrm{se}} n_{\nu} \right) [\mathrm{E^\ast}] \, ,
\end{align}
\end{subequations}
denote the directed fluxes and $n_{\nu}$ the molar concentration of photons of frequency $\nu$ or angular frequency $\omega_\nu = 2\pi \nu$.

In addition to the photophysical transitions, we also consider a nonradiative thermal pathway connecting the ground and the excited states, as depicted in Figure~\ref{FIG_ab:em_elementary}c,
\begin{equation}
    \ce{E{^\ast} <=>[k_{+\mathrm{q}}][k_{-\mathrm{q}}] E} \, .
    \label{eq:chemEL}
\end{equation}
This is usually called \emph{quenching} reaction and we assume it to be elementary.
This implies that its current follows mass action kinetics
\begin{equation}
    J_\mathrm{q} = J_{+\mathrm{q}} - J_{-\mathrm{q}} \, ,
    \label{eq:chemCurr}
\end{equation}
with
\begin{subequations}
\begin{align}
    	J_{+\mathrm{q}} & = k_{+\mathrm{q}} [\mathrm{E^\ast}] \\
    	J_{-\mathrm{q}} & = k_{-\mathrm{q}} [\mathrm{E}] \, .
\end{align}
\end{subequations}

Overall, the dynamics of the elementary photophysical and quenching reactions described in terms of concentrations reads
\begin{subequations}
\begin{align}
	\dt [\mathrm{E^\ast}] =& - \dt [\mathrm{E}] = J_\nu - J_\mathrm{q} \, , \label{eq:rates_a} \\
	\dt n_\nu =& - J_\nu \, . \label{eq:rates_b}
\end{align}
\label{eq:rates}
\end{subequations}
As a consequence, the total concentration of $\mathrm{E}$-molecules, denoted $\mathrm{E}_0 \equiv [\mathrm{E}] + [\mathrm{E^\ast}]$, is conserved
\begin{equation}
\dt \mathrm{E}_0 = \dt ([\mathrm{E}] + [\mathrm{E^\ast}]) = 0 \, ,
\label{eq:LE}
\end{equation}
but not the concentration of photons $n_\nu$.

\subsection{Thermodynamics}
\label{sec:thermopot}

Nonequilibrium thermodynamics of chemical reactions crucially relies on the \textit{local equilibrium} assumption. 
This means that all degrees of freedom other than those involved in reactions (\textit{i.e.} concentrations) are considered to be at equilibrium~\cite{prigogine1949,Kondepudi2014}.
As a result, \textit{(i)} the temperature $T$ is well defined and it is set by the solvent, which acts as a thermal reservoir, and \textit{(ii)} the chemical potentials of all chemical species are evaluated using their equilibrium expression in solution (expressed as energy minus temperature multiplying entropy) but evaluated at the nonequilibrium values of the concentrations~\cite{Kondepudi2014}.

By focusing first on the chemical species, we recall that, for an ideal dilute and homogeneous solution, the total internal energy $U_{\mathrm{ch}}$ and entropy $S_{\mathrm{ch}}$ per unit volume read \cite{rao2016}:
\begin{align}
	U_{\mathrm{ch}} & = u^\circ_{\mathrm{E}} [\mathrm{E}] + u^\circ_\mathrm{E^\ast} [\mathrm{E^\ast}] \label{eq:Ech} \\
	& = u^\circ_{\mathrm{E}} [\mathrm{E}] + (\NA \hbar \omega_\nu + u^\circ_\mathrm{E}) [\mathrm{E^\ast}] \notag\\
	S_{\mathrm{ch}} & =  ( s_{\mathrm{E}} + R ) [\mathrm{E}] + ( s_{\mathrm{E^\ast}} + R ) [\mathrm{E^\ast}]  \label{eq:Sch} \, ,
\end{align}
where
\begin{align}
    s_\mathrm{E} &:= s^\circ_\mathrm{E} - R \ln [\mathrm{E}] \, , &
    s_\mathrm{E^\ast} &:= s^\circ_\mathrm{E} - R \ln [\mathrm{E^\ast}] \, ,
\end{align}
denote the molar entropy of formation carried by the chemical species in the solution.
Also, $R$ denotes the gas constant while $\NA$ the Avogadro's number.
Note that we assumed that the energy difference between the ground and the excited state is exactly the energy carried by a photon of frequency $\nu$, \textit{i.e.} $u^\circ_\mathrm{E\ast} - u^\circ_\mathrm{E} = \NA \hbar\omega_\nu$.
In Eq.~\eqref{eq:Sch}, the $s^\circ$ terms take into account the possible degeneracy of the ground states, usually denoted $g$ (\textit{i.e.}, $s^\circ = R \ln g$).
The total free energy per unit volume of the chemical species consequently reads
\begin{equation}
F_{\mathrm{ch}} = U_{\mathrm{ch}} - T  S_{\mathrm{ch}}  \, .
\end{equation}
From this expression, the chemical potentials ensue,
\begin{subequations}
\begin{align}
	\mu_\mathrm{E} = \partial_{[\mathrm{E}]} F_{\mathrm{ch}} & = u^\circ_\mathrm{E} - Ts_\mathrm{E} \\
	& = \mu^\circ_\mathrm{E} + RT \ln [\mathrm{E}] \\
	\mu_\mathrm{E^\ast} = \partial_{[\mathrm{E^\ast}]} F_{\mathrm{ch}} & = u^\circ_\mathrm{E^\ast} - Ts_\mathrm{E^\ast} \\ & = \mu^\circ_\mathrm{E^\ast} + RT \ln [\mathrm{E^\ast}] \, ,
\end{align}
\label{eq:muchem}
\end{subequations}
where the standard chemical potentials are given by
\begin{subequations}
    \begin{align}
        &\mu^\circ_\mathrm{E} = u^\circ_\mathrm{E} - T s^\circ_\mathrm{E} \, , &
        &\mu^\circ_\mathrm{E^\ast} = u^\circ_\mathrm{E^\ast} - T s^\circ_\mathrm{E^\ast} \, .
    \end{align}
\end{subequations}

We now extend this local equilibrium reasoning to the radiation. 
We thus introduce the equilibrium expressions for the thermodynamic potentials of radiation, but evaluate them at arbitrary photons' concentration.
The energy per unit volume and frequency carried by photons thus reads 	
\begin{equation}
    U_{\mathrm{ph}} = \NA \hbar \omega_\nu n_\nu \, , \label{IntEnergPh}
\end{equation}
and the entropy per unit volume and frequency (obtained using the expression derived for equilibrium Bose gasses~\cite{Kondepudi2014}) reads
\begin{align}
	\hspace{-1em}
        S_{\mathrm{ph}} =  R \bigg[ \left(f_\nu +  n_\nu\right) \ln \left(f_\nu +  n_\nu\right) - n_\nu \ln n_\nu - f_\nu \ln f_\nu  \bigg] \, . 
	\label{eq:Sph}
\end{align}
Note that this local equilibrium assumption is equivalent to neglecting eigentstates coherences from the von Neumann entropy of the radiation. 
As further discussed in Sec.~\ref{sec:discussion}, it implies that the radiation may be out-of-equilibrium with respect to the thermal reservoir but incoherent.
For large photon numbers~---~the limit of interest for this paper~---~the corresponding molar quantities are obtained as derivatives with respect to $n_\nu$:
	\begin{align}
	    u_{\nu} & = \partial_{n_\nu} U_{\mathrm{ph}} = \NA \hbar \omega_\nu \label{eq:unu} \\
		s_{\nu} & = \partial_{n_\nu} S_{\mathrm{ph}} = R \ln\left\{ \left( f_\nu +  n_{\nu} \right)/ n_{\nu} \right\} \, .
		\label{eq:snu} 
	\end{align}
The energy of a photon is an intrinsic property which only depends on its frequency, while for chemical species the same quantity must be specified with respect to some reference condition (symbol $^\circ$ is used to indicate standard temperature and pressure).

In analogy with Eq.~\eqref{eq:muchem}, we introduce the chemical potential of a mole of photons with frequency $\nu$ with respect to the temperature $T$ of the solvent
\begin{equation}
    \mu_{\nu} =  u_{\nu} - T s_{\nu} = \NA \hbar \omega_\nu - RT \ln \frac{ f_\nu +  n_{\nu} }{ n_{\nu} } \, .
    \label{eq:munu} 
\end{equation}
It measures the free energy carried by the photons with respect to the molecules in solution.

For \emph{thermal radiation} at temperature $T_\mathrm{r}$, we get that 
\begin{equation}
   \mu^\mathrm{th}_\nu = u_\nu \left(1 - \frac{T}{T_\mathrm{r}} \right) \, ,
   \label{eq:mu_thermal}
\end{equation}
since in that case~\cite{Kondepudi2014,meszena1999}:
\begin{equation}
    n^{\mathrm{th}}_\nu = \frac{f_\nu \langle n_\nu \rangle}{\NA} \, ,
    \label{eq:plank}
\end{equation}
where
\begin{equation}
    f_\nu = \frac{2 \omega_\nu^2}{\pi c^3} \quad \text{and} \quad \langle n_\nu \rangle = \frac{1}{\mathrm{e}^{\NA\hbar \omega_\nu / R T_\mathrm{r}} - 1}
\end{equation}
denote the density of photon states associated with the radiation and the Bose--Einstein distribution, respectively.
When the temperature of radiation and solvent coincide, $T_\mathrm{r} = T$, the chemical potential of the photons vanishes, in line with the fact that photons are not conserved in physical systems in contact with a thermal reservoir~\cite{wurfel2005}.
Instead, in the limit of $T_\mathrm{r} \rightarrow \infty$ the chemical potential of thermal photons coincides with the energy they carry.
We note that the expression \eqref{eq:mu_thermal} for chemical potential of thermal photons is consistent with previous formulations \cite{ries1991,Milazzo1997,meszena1999,chen2008,Wurfel2016}.
Differently from these, our formulation explicitly leverages on the condition of local-equilibrium, in analogy with the approach of nonequilibrium thermodynamics of chemical reactions~\cite{rao2016,avanzini2021ni}.

\subsection{Local Detailed Balance}
\label{sec:ldb}

The central property connecting kinetic rate constants and thermodynamic chemical potentials is the \emph{local detailed balance}, also referred as \emph{microscopic reversbility}~\cite{blackmond2009,astumian2012,astumian2018}.
Using this property one can relate the thermodynamic forces driving each reaction, also called reaction \emph{affinities}, to the corresponding forward and backward fluxes.

For the quenching reaction~\eqref{eq:chemEL}, the local detailed balance property reads
\begin{equation}
\frac{k_{+\mathrm{q}}}{k_{-\mathrm{q}}} = \exp\left\{ \frac{\mu^\circ_\mathrm{E^\ast} - \mu^\circ_\mathrm{E}}{RT} \right\} \, ,
\label{eq:ldl_constants}
\end{equation}
and, using \eqref{eq:chemCurr} and \eqref{eq:muchem}, it enables us to identify the quenching affinity as
\begin{equation}
    A_\mathrm{q} \equiv  \mu_\mathrm{E^\ast} - \mu_\mathrm{E} = RT \ln \frac{J_{+\mathrm{q}}}{J_{-\mathrm{q}}} \, .
    \label{eq:ldb_thermal}
\end{equation}
This last relation binds the affinity to the reaction fluxes, and implies that chemical forces and reaction currents are always aligned.
We emphasize that Eqs.~\eqref{eq:ldl_constants} and~\eqref{eq:ldb_thermal} hold for elementary reactions but not necessarily for coarse-grained processes, as we will see (see also Refs.~\cite{wachtel2018,avanzini2020cg}).

For the photophysical reaction, the local detailed balance is essentially the Einstein relations~\cite{einstein1916}
\begin{subequations}
	\begin{align}
		\frac{k_{\mathrm{a}}}{k_{\mathrm{se}}} & = \exp\left\{ \frac{s^\circ_{\mathrm{E^\ast}} - s^\circ_{\mathrm{E}}}{R} \right\} \label{eq:einstein1} \\
		\frac{k_{\mathrm{e}}}{k_{\mathrm{se}}} & = f_\nu \label{eq:einstein2} \, .
	\end{align}
	\label{eq:einstein}
\end{subequations}
Substituting them into Eq.~\eqref{eq:ab--emCurrent}, we can express the forward and backward currents as
\begin{subequations}
\begin{align}
    J_{+\nu} &= k_\mathrm{a} n_\nu [\mathrm{E}] = \e^{(s^\circ_{\mathrm{E^\ast}} - s^\circ_{\mathrm{E}}) / R} k_\mathrm{se} n_\nu [\mathrm{E}]  \label{eq:forwardphotoflux} \\ 
    J_{-\nu} &= (k_\mathrm{e} + k_\mathrm{se} n_\nu) [\mathrm{E^\ast}] = k_\mathrm{se} (f_\nu + n_\nu) [\mathrm{E^\ast}] \, , \label{eq:backwardphotoflux}
\end{align}
    \label{eq:photofluxes}
\end{subequations}
and using Eqs.~\eqref{eq:muchem}, \eqref{eq:snu}, we find
\begin{equation}
A_\nu \equiv \mu_{\mathrm{E}} + \mu_{\nu} - \mu_{\mathrm{E^\ast}} = RT \ln \frac{J_{+\nu}}{J_{-\nu}} \, ,
	\label{eq:ldb_photo}
\end{equation}
where $A_\nu$ is the affinity of the photophysical reaction.

From a thermodynamic standpoint, \emph{equilibrium} is reached when all affinities vanish
\begin{subequations}
\begin{align}
    &A^\mathrm{eq}_\mathrm{q} = \mu^\mathrm{eq}_\mathrm{E^\ast} - \mu^\mathrm{eq}_\mathrm{E} = 0 \\
    &A^\mathrm{eq}_\nu = \mu^\mathrm{eq}_{\mathrm{E}} + \mu^\mathrm{eq}_{\nu} - \mu^\mathrm{eq}_{\mathrm{E^\ast}} = 0 \;.
\end{align}
\label{eq:equilibrium_thermo}
\end{subequations}
Therefore, the chemical potential of the photons relative to the solution must vanish, $\mu^\mathrm{eq}_{\nu} = 0$.
This obviously implies that the temperature of the radiation and the temperature of the thermal reservoir must be equal, see Eq.~\eqref{eq:mu_thermal}.
Finally, through Eqs.~\eqref{eq:ldb_thermal} and~\eqref{eq:ldb_photo}, Eq.~\eqref{eq:equilibrium_thermo} implies that the equilibrium currents vanish as well:
\begin{equation}
    J^\mathrm{eq}_\nu = 0 \, , \quad \text{and} \quad \quad J^\mathrm{eq}_\mathrm{q} = 0 \, .
    \label{eq:equilibrium_dyn}
\end{equation}

\subsection{Molecules and radiation in a closed box}
\label{sec:closedsystem}

\begin{figure}[tb]
    \includegraphics[width=.35\textwidth]{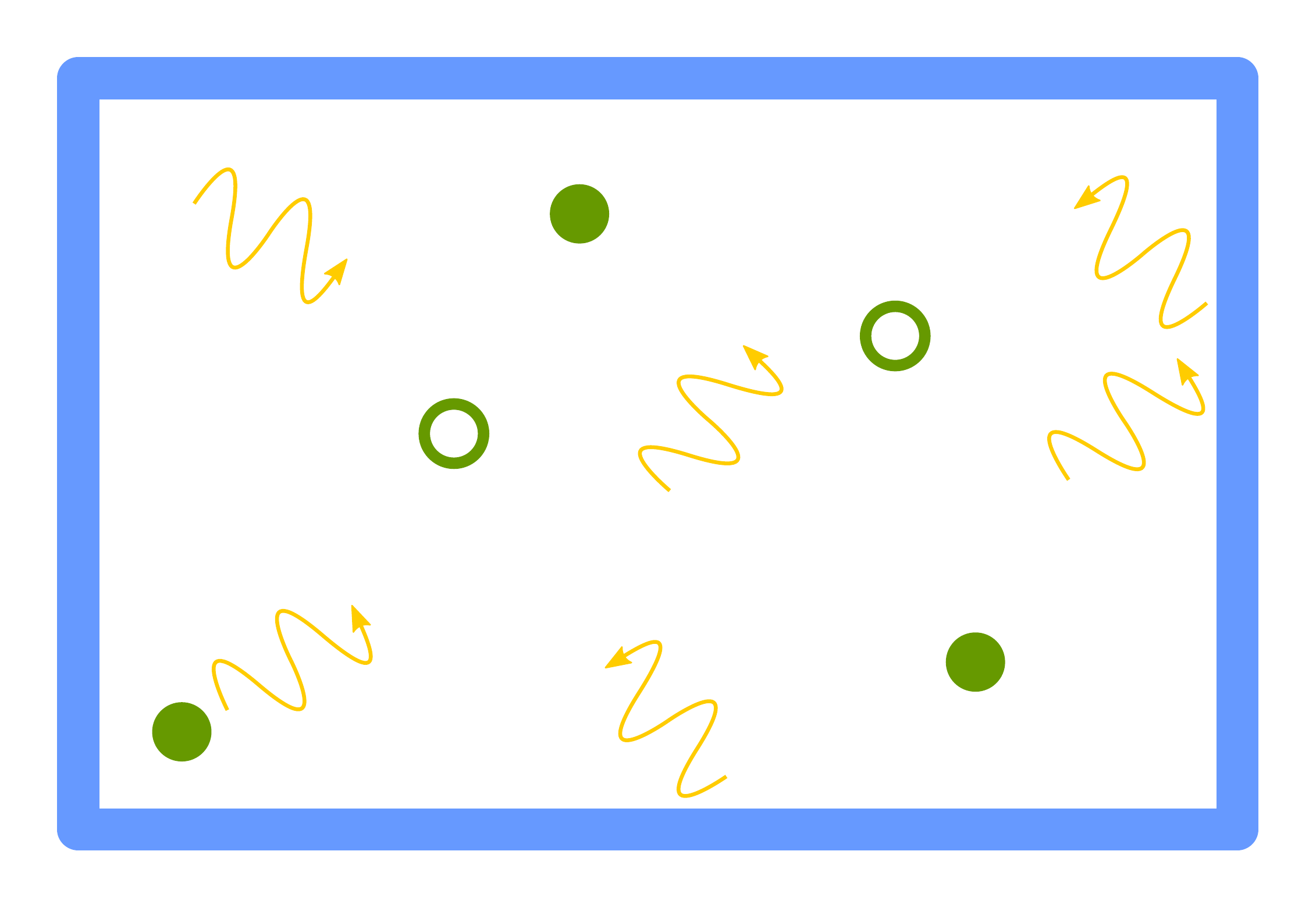}
    \centering
    \caption{\textbf{| Closed Box.}
    Radiation interacting with species $\mathrm{E}$ and $\mathrm{E^\ast}$ via the scheme of Figure~\ref{FIG_ab:em_elementary} in an ideal dilute solution at temperature $T$ inside a closed box with perfectly reflecting walls. Photons cannot leave the system but their number is not conserved due to the coupling between light-induced and quenching reactions.
}
\label{FIG_photophysics2} 
\end{figure}

We are now ready to consider the situation where the radiation and the molecules in solution at temperature $T$ are enclosed in a box with perfectly reflecting internal walls, as depicted in Figure~\ref{FIG_photophysics2}.
The system is thus closed with respect to both matter and photons. 
Its state is defined by the concentrations of photons and chemical species $n_\nu$, $[\mathrm{E}]$ and $[\mathrm{E^\ast}]$, which obey the nonlinear dynamics~\eqref{eq:rates}.
The condition of steady state implies equilibrium between the photophysical and quenching reactions, see Eq.~\eqref{eq:equilibrium_dyn}.
This also implies that
\begin{equation}
    \frac{[\mathrm{E^\ast}]^\mathrm{eq}}{[\mathrm{E}]^\mathrm{eq}} = \frac{k_{-\mathrm{q}}}{k_{+\mathrm{q}}} = \frac{k_{\mathrm{a}} n^\mathrm{eq}_{\nu}}{k_{\mathrm{e}} + k_{\mathrm{se}} n^\mathrm{eq}_{\nu}} \, ,
    \label{eq:dynEQ}
\end{equation}
and more explicitly that 
\begin{equation}
\hspace{-2em}
[\mathrm{E}]^\mathrm{eq} = \mathrm{E}_0 - [\mathrm{E^\ast}]^\mathrm{eq} 
= \frac{k_{+\mathrm{q}} \mathrm{E}_0}{k_{+\mathrm{q}}+ k_{-\mathrm{q}}}
= \frac{(k_{\mathrm{e}} + k_{\mathrm{se}} n_{\nu}^\mathrm{eq}) \mathrm{E}_0}{k_{\mathrm{e}} +  k_{\mathrm{se}} n_{\nu}^\mathrm{eq} + k_{\mathrm{a}} n_{\nu}^\mathrm{eq}} \;,
\end{equation}
where the total concentration of chemicals $\mathrm{E}_0$ is set by the initial condition.

Turning to energetic considerations, the first law of thermodynamics expresses the fact that the change in internal energy of the system per unit time ---~evaluated using \eqref{eq:Ech} and \eqref{IntEnergPh} 	with \eqref{eq:rates}~--- is due to the heat flow $\dot{Q}$ entering the system
\begin{align}
    d_t \left( U_{\mathrm{ch}} + U_{\mathrm{ph}}\right) = - \NA \hbar\omega_\nu J_\mathrm{q} = \dot{Q}\; .
    \label{eq:closed_Ilaw}
\end{align}
The second law states that the entropy production ---~defined as the entropy change in the system and in the reservoir (\textit{viz.} the solvent)~--- must always be greater or equal than zero. This is verified as 
\begin{align}
    \dot{\Sigma} = d_t \left( S_{\mathrm{ch}} + S_{\mathrm{ph}}\right) - \frac{\dot{Q}}{T} = J_\nu \frac{A_\nu}{T} + J_\mathrm{q} \frac{A_\mathrm{q}}{T} \ge 0 \, ,
        \label{eq:closed_IIlaw}
\end{align}
where we used \eqref{eq:Sch} and \eqref{eq:Sph} with \eqref{eq:rates} and the affinities \eqref{eq:ldb_thermal} and \eqref{eq:ldb_photo}.
In particular, the inequality readily follows from Eq.~\eqref{eq:ldb_thermal}.

We now show that this system relaxes to equilibrium and its relaxation is well characterized by the nonequilibrium total free energy
\begin{equation}
     F = F_\mathrm{ch} + F_{\mathrm{ph}} \, ,
     \label{eq:noneqF}
\end{equation}
with
\begin{subequations}
\begin{align}
    &F_\mathrm{ch} = U_\mathrm{ch} - TS_\mathrm{ch} = (\mu_\mathrm{A} - R) [\mathrm{E}] + (\mu_\mathrm{E^\ast} - R) [\mathrm{E^\ast}] \\
    &F_{\mathrm{ph}} = U_{\mathrm{ph}} - TS_{\mathrm{ph}} = \mu_\nu n_\nu -  RT f_\nu \ln (f_\nu + n_\nu) \, ,
\end{align}
\end{subequations}
where the temperature $T$ of the solution has been used as reference in the expression of both $F_\mathrm{ch}$ and $F_{\mathrm{ph}}$.
Indeed, from Eqs.~\eqref{eq:closed_Ilaw} and \eqref{eq:closed_IIlaw} this total free energy can only decrease:
\begin{equation}
    \dt F = \dt (F_\mathrm{ch} + F_{\mathrm{ph}}) = - T\dot{\Sigma} \le 0 \;.
    \label{eq:lyapounov1}
\end{equation}
Also, the total free energy is lower bounded by its equilibrium value:
\begin{align}
    &\hspace{-0.4cm}(F_\mathrm{ch} + F_{\mathrm{ph}}) - (F^\mathrm{eq}_\mathrm{ch} + F^\mathrm{eq}_{\mathrm{ph}}) = \label{eq:lyapounov2} \\
    &\hspace{0.3cm} [\mathrm{E^\ast}] RT\ln \frac{[\mathrm{E^\ast}]}{[\mathrm{E^\ast}]_\mathrm{eq}} - RT([\mathrm{E^\ast}] - [\mathrm{E^\ast}]_\mathrm{eq}) \nonumber \\
    &\hspace{0.3cm} +[\mathrm{E}] RT\ln \frac{[\mathrm{E}]}{\mathrm{[\mathrm{E}]_\mathrm{eq}}} - RT([\mathrm{E}] -\mathrm{[\mathrm{E}]_\mathrm{eq}})  \nonumber \\
    &\hspace{0.3cm} +f_{\nu} RT \ln \frac{f_{\nu} + n^\mathrm{eq}_{\nu}}{f_{\nu}+ n_{\nu}} + n_{\nu} RT \ln \frac{ n_{\nu} \left(f_{\nu} + n^\mathrm{eq}_{\nu} \right)}{n^\mathrm{eq}_{\nu} \left(f_{\nu} + n_{\nu} \right)}\ge 0 \, . \nonumber
\end{align}
The first equality follows from
\begin{equation}
([\mathrm{E^\ast}]_\mathrm{eq} - [\mathrm{E^\ast}]) \mu_\mathrm{E^\ast}^\mathrm{eq} = ([\mathrm{E}]_\mathrm{eq} - [\mathrm{E}]) \mu_\mathrm{E}^\mathrm{eq} + (n_\nu^\mathrm{eq} - n_\nu) \mu_\nu^\mathrm{eq} \, ,
\end{equation}
which holds at any time by virtue of the equilibrium condition~\eqref{eq:equilibrium_thermo} and the conservation law~\eqref{eq:LE}.
The inequality is set by the properties of logarithms (log inequality).
Overall, Eqs.~\eqref{eq:lyapounov1} and \eqref{eq:lyapounov2} show that the total free energy acts as a Lyapunov function:
for any initial condition, the radiation will eventually thermalize with the solution by minimizing $F_\mathrm{ch} + F_{\mathrm{ph}}$.

We note that this treatment can be straightforwardly extended to a multi-frequency situation by including more chemical species or more photophysical and quenching transitions in the model.

\subsection{Molecules and radiation in a transparent box}
\label{sec:radiostatted}

\begin{figure}[h!]
    \includegraphics[width=.45\textwidth]{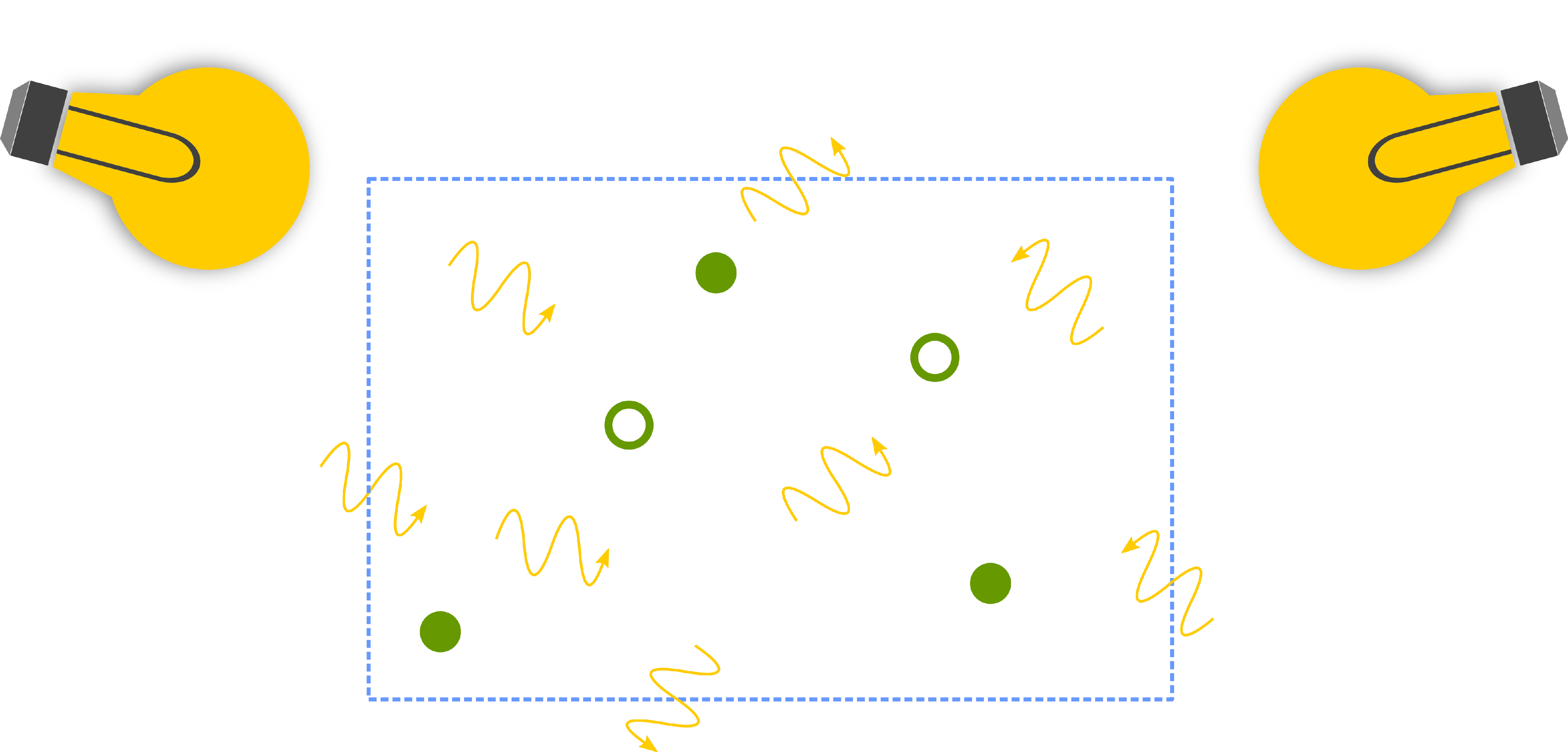}
    \centering
    \caption{\textbf{| Transparent box.}
    Radiation interacting with species $\mathrm{E}$ and $\mathrm{E^\ast}$ via the scheme of Figure~\ref{FIG_ab:em_elementary} in an ideal dilute solution at temperature $T$ inside a box with perfectly transparent walls. The system is closed to exchanges of molecules but is open to photons, which can freely enter and exit the system. Photons' concentration is controlled by a radiostat. 
}
\label{FIG_photophysics4} 
\end{figure}

We now turn to the situation where an external mechanism keeps the concentration of photons $n_\nu$ constant inside the system, as illustrated in Figure~\ref{FIG_photophysics4}.
We will refer to such mechanism as \emph{radiostat}, in analogy with \emph{thermostats} and \emph{chemostats}.
As $n_\nu$ is not anymore a dynamical variable, only equation~\eqref{eq:rates_a} defines the dynamics, which is now linear and can be easily solved analytically. Equation~\eqref{eq:rates_b} only defines the rate at which photons must enter the system to keep their concentration constant.   
The steady-state condition will now generically produce nonvanishing currents $\bar{J}_\nu =\bar{J}_\mathrm{q}$ corresponding to a nonequilibrium steady state:
\begin{equation}
[\overline{\mathrm{E}}] = \mathrm{E}_0 - [\overline{\mathrm{E^\ast}}] = \frac{(k_{+\mathrm{q}} + k_{\mathrm{e}} + k_{\mathrm{se}} n_{\nu}) \mathrm{E}_0}{k_{+\mathrm{q}}+ k_{-\mathrm{q}} + k_{\mathrm{e}} + k_{\mathrm{se}} n_{\nu} + k_{\mathrm{a}} n_{\nu}} \;.
\end{equation}
From now on, overbars will denote quantities at steady state.

We now proceed with thermodynamics.
As the radiation inside the box is held constant by the radiostat, the only contribution to the change in the system's internal energy per unit time comes from the molecules in solution and can be decomposed ---~using~\eqref{eq:Ech} with \eqref{eq:rates}, \eqref{eq:ab--emCurrent} and~\eqref{eq:chemCurr}~--- as
\begin{equation}
    d_t U_\mathrm{ch} = \dot{U}_\nu + \dot{Q} \, , 
    \label{eq:transparent_dtU}
\end{equation}
where
\begin{equation}
    \dot{Q}=- \NA \hbar\omega_\nu J_\mathrm{q}
\end{equation}
is the rate of heat absorbed by the molecules, and
\begin{equation}
    \dot{U}_\nu = J_\nu u_\nu
\end{equation}
is the rate of energy absorbed by the molecules from the photons.
The entropy production rate now reads
\begin{equation}
   \dot{\Sigma} 
   = d_t S_\mathrm{ch} - \frac{\dot{Q}}{T} - \dot{S}_\nu 
   = J_\nu \frac{A_\nu}{T} + J_\mathrm{q} \frac{A_\mathrm{q}}{T} \ge 0 \, .
    \label{eq:transparent_dtS}
\end{equation}
In contrast to Eq.~\eqref{eq:closed_IIlaw}, $\dot{\Sigma}$ now accounts for the rate of entropy exchanged with the radiostat, $\dot{S}_\nu = J_\nu s_\nu$.
To obtain the second equality ensuring the non-negativity of the entropy production, we used \eqref{eq:Sch} with \eqref{eq:rates} and the affinities \eqref{eq:ldb_thermal} and \eqref{eq:ldb_photo}.

By combining eqs.~\eqref{eq:transparent_dtU} and \eqref{eq:transparent_dtS}, we get the free-energy balance 
\begin{align}
    T\dot{\Sigma} 
    = \dot{F}_\nu - d_t F_\mathrm{ch} \ge 0 \, ,
    \label{eq:EP_TransClosed_Kelvinlike}
\end{align}
where
\begin{equation}
    \dot{F}_\nu = J_\nu \mu_\nu
    \label{eq:Fdotnu}
\end{equation}
represents the rate of free energy absorbed by the molecules in solution from the radiation.
This result shows that the variation of free energy in the system (\textit{e.g.} the free energy stored) is upper bounded by the free energy absorbed from the photons.

We consider now the special case where the radiostat is a black body emitting thermal radiation  at temperature $T_\mathrm{r}$ (\textit{e.g.} a lamp).
From Eqs.~\eqref{eq:snu} and~\eqref{eq:plank}, and as expected from a thermal object, $\dot{U}_\nu = T_\mathrm{r} \dot{S}_\nu$.
As a consequence, the term describing radiation absorption in the energy balance ($\dot{U}_\nu$) enters the entropy balance as energy over the radiation temperature ($\dot{S}_\nu = \dot{U}_\nu / T_\mathrm{r}$).
In other words, the radiostat acts as a thermal reservoir ---~at temperature $T_\mathrm{r}$~--- that exchanges heat with the systems through photons.
The free energy carried by photons is given by the chemical potential $\mu^\mathrm{th}_\nu$, Eq.~\eqref{eq:mu_thermal}.
Hence, the rate of radiation free energy entering the system, Eq.~\eqref{eq:Fdotnu}, becomes
\begin{equation}
    \dot{F}_\nu = J_\nu \mu^\mathrm{th}_\nu = \dot{U}_\nu \left(1 - \frac{T}{T_\mathrm{r}} \right) \, .
    \label{eq:heatbb}
\end{equation}
When $T_\mathrm{r} = T$, we have $\dot{F}_\nu = 0$ and the system relaxes to equilibrium by minimizing $F_\mathrm{ch}$, as in the closed case, see Eq.~\eqref{eq:EP_TransClosed_Kelvinlike}.
But whenever there is a mismatch between the temperature of the radiation reservoir and the temperature of the thermal reservoir, the system will be driven out of equilibrium, and eventually reach a nonequilibrium steady state.
At steady state, if $T_\mathrm{r}> T$, the heat flows from the radiation to the solvent, and vice versa if $T_\mathrm{r}< T$.
If one starts from an equilibrium solution at temperature $T$ and turns on the radiation temperature $T_\mathrm{r}> T$, 
the radiation heat flow can be used as a resource to drive and sustain accumulation of free energy in the system. 
Such energy storage phenomena is reminiscent of what happens in photosynthetic systems, as sunlight can be regarded as black body radiation at a temperature of about $5800$ Kelvin.

When the temperature of the black body radiation becomes very large,  $T_\mathrm{r} \to \infty$, (more realistically when $\NA \hbar \omega_\nu \ll R T_\mathrm{r}$) spontaneous emission becomes negligible compared to absorption and stimulated emission, and the entropy of the radiation vanishes: $s_\nu \rightarrow 0$. 
As a consequence, the free energy absorbed is exclusively made of energy  
\begin{equation}
    \dot{F}_\nu = J_\nu \mu_\nu \rightarrow J_\nu u_\nu = \dot{U}_\nu \, ,
    \label{eq:laserWork}
\end{equation}
and the radiation can be regarded as work source.


\section{Basic photochemical mechanisms}
\label{sec:photochemistry}

So far, we neglected the possibility that additional excited-state reactive events convert the species $\mathrm{E^\ast}$ into a different one, $\mathrm{Z}$.
The net effect, $\mathrm{E} \ce{->[\text{light}]} \mathrm{Z}$, is in fact the crux of photochemistry. 
Often, the details of such excited-state dynamics are not explicitly specified in the kinetic descriptions~\cite{feringa2009,sabatino2019}.
However, as we will see in this section, distinct mechanisms lead to different thermodynamics.
We will consider two of the most common models for photochemical unimolecular reactions, namely the adiabatic mechanism (\S~\ref{sec:adiabatic}) and the diabatic one (\S~\ref{sec:diabatic})~\cite{foerster1970,Balzani2014}.
To connect our results to experimental observations, we derive the coarse-grained description of both mechanisms in \S~\ref{sec:cg}.
Such description does not require a detailed knowledge of excited state dynamics.

\subsection{Adiabatic mechanism}
\label{sec:adiabatic}

\begin{figure}[tb]
    \includegraphics[width=.50\textwidth]{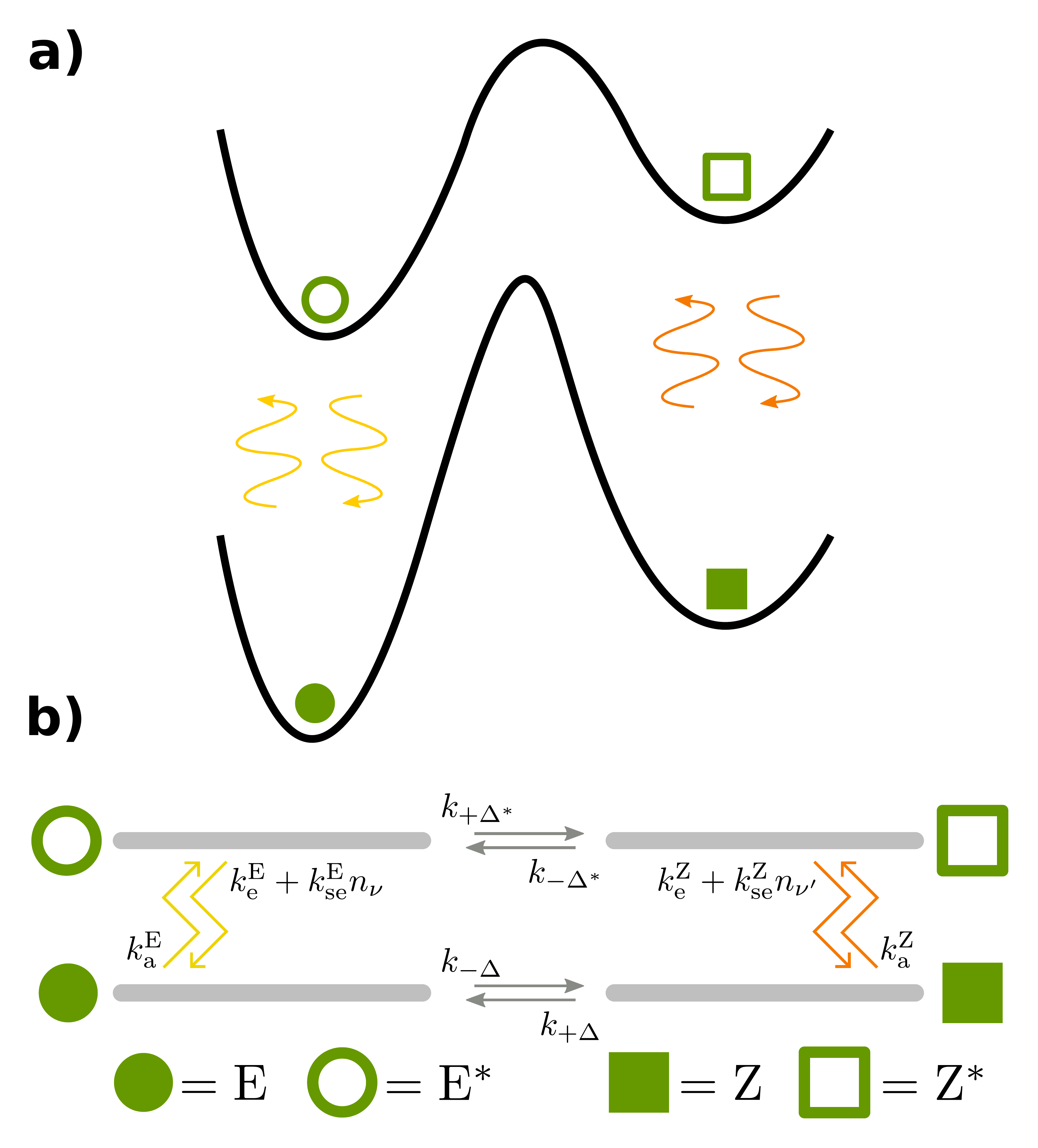}
    \centering
    \caption{\textbf{| Adiabatic mechanism.}
    \textbf{a}) Pictorial representation of the ground and excited state potential energy surfaces along the reaction coordinate interconverting species $\ce{E}$ and $\ce{Z}$.
    \textbf{b}) Schematic representation (also known as Jablonski diagram~\cite{Balzani2014}) of the mechanism.
    This scheme can represent F{\"o}ster cycles describing excited-state proton transfer reactions, where for instance species $\mathrm{Z}$ is a zwitterionic tautomer~\cite{iupac1994} of $\mathrm{E}$.~\cite{grabowski1977,formosinho1993, Balzani2014}
}
\label{FIG_photochemistry_pes} 
\end{figure}

Consider the photoisomerization scheme in Figure~\ref{FIG_photochemistry_pes}.
It comprises two elementary photophysical reactions,
\begin{align}
    \mathrm{E} &\ce{<=>[\nu]} \mathrm{E}^{\ast} \, , \quad \text{and } & \mathrm{Z} &\ce{<=>[\nu']} \mathrm{Z}^{\ast} \, ,
    \label{eq:adiabaticPP}
\end{align}
and two thermally-induced non-radiative reactions,
\begin{align}
    \mathrm{E} &\ce{<=>[\Delta]} \mathrm{Z} \, , \quad \text{and } & \mathrm{E}^{\ast} &\ce{<=>[\Delta^\ast]} \mathrm{Z}^{\ast} \, ,
\end{align}
which connect species in the same electronic state.
Note that \textit{(i)} the two photophysical reactions are activated by photons of different frequencies, $\nu$ and $\nu'$, and \textit{(ii)} we disregard the quenching reactions associated to the photophysical reactions \eqref{eq:adiabaticPP}.

For the moment, we consider the concentrations of photons $n_\nu$ and $n_{\nu'}$ as controlled by two different radiostats (see discussion in \S~\ref{sec:radiostatted}).
The system's dynamics is ruled by the net absorption current of photons $\gamma_\nu$ going from $\mathrm{E}$ to $\mathrm{E^\ast}$ and the one of photons $\gamma_{\nu'}$ going from $\mathrm{Z}$ to $\mathrm{Z^\ast}$,
\begin{subequations}
    \begin{align}
        J_{\nu} &= k_\mathrm{a}^\mathrm{E} n_\nu [\mathrm{E}] - (k_\mathrm{e}^\mathrm{E} + k_\mathrm{se}^\mathrm{E} n_\nu) [\mathrm{E^\ast}] \\
        J_{\nu'}  &=  k_\mathrm{a}^\mathrm{Z} n_{\nu'} [\mathrm{Z}] - (k_\mathrm{e}^\mathrm{Z} + k_\mathrm{se}^\mathrm{Z} n_{\nu'}) [\mathrm{Z^\ast}] \, ,
    \end{align}
\label{eq:photocurrents}
\end{subequations}
and by the reaction currents from $\mathrm{E^\ast}$ to $\mathrm{Z^\ast}$ and from $\mathrm{Z}$ to $\mathrm{E}$,
\begin{subequations}
    \begin{align}
        J_{\Delta^\ast} &= k_{+\Delta^\ast} [\mathrm{E^\ast}] - k_{-\Delta^\ast} [Z^\ast] \\
        J_{\Delta} &= k_{+\Delta} [\mathrm{Z}] - k_{-\Delta} [\mathrm{E}] \label{eq:ADthermocurrent} \, .
    \end{align}
\label{eq:photochemistry_stardynamics}
\end{subequations}
It is easy to realize that, at the steady state, the currents must satisfy the following relations:
\begin{equation}
    \bar{J} \equiv \bar{J}_\nu = \bar{J}_{\Delta^\ast} = - \bar{J}_{\nu'} =  \bar{J}_\Delta \, ,
    \label{eq:DissADss}
\end{equation}
with the special case $\bar{J} = 0$ describing equilibrium.

The total internal energy and entropy of the system are given by the expressions in Eqs.~\eqref{eq:Ech} and~\eqref{eq:Sch}, respectively, but summed over all chemical species:
\begin{align}
    U_\mathrm{ch} = \sum_{\substack{X=\mathrm{E},\\\mathrm{E^\ast},\mathrm{Z},\mathrm{Z^\ast}}} u^\circ_\mathrm{X} [\mathrm{X}] \, , \quad
    S_\mathrm{ch} = \sum_{\substack{X=\mathrm{E},\\\mathrm{E^\ast},\mathrm{Z},\mathrm{Z^\ast}}} (s_\mathrm{X} + R) [\mathrm{X}] \, .
	\label{eq:totalEch}
\end{align}
The total free energy is thus given by $F_\mathrm{ch} = U_\mathrm{ch} - T S_\mathrm{ch}$.
Based on the results from the previous section, Eq.~\eqref{eq:transparent_dtS}, the entropy production can be expressed as the sum of currents times affinities,
\begin{align}
    \dot{\Sigma} = \sum_{\substack{\rho=\nu,\\{\nu'},\Delta,\Delta^\ast}} J_\rho \frac{A_\rho}{T} =
    R \sum_{\substack{\rho=\nu,\\{\nu'},\Delta,\Delta^\ast}} J_\rho \ln \frac{J_{+\rho}}{J_{-\rho}} \ge 0 ,
    \label{eq:EP_photoADfluxforce}
\end{align}
where the sum runs over all the reactions, Eqs.~\eqref{eq:photocurrents} and~\eqref{eq:photochemistry_stardynamics}.
As before (Eq.~\eqref{eq:EP_TransClosed_Kelvinlike}), by combining Eqs.~\eqref{eq:EP_photoADfluxforce} and~\eqref{eq:totalEch} we get the free-energy balance as
\begin{equation}
     T\dot{\Sigma} = -d_t F_\mathrm{ch} + \dot{F}_\nu + \dot{F}_{\nu'} \, ,
    \label{eq:EP_photochemistryAD}
\end{equation}
where the free-energy absorbed has contributions from both the light sources: $\dot{F}_\nu = J_\nu \mu_\nu$ and $\dot{F}_{\nu'} =  J_{\nu'} \mu_{\nu'}$.

We now focus to steady states.
Using Eq.~\eqref{eq:DissADss}, we can rewrite the entropy production \eqref{eq:EP_photochemistryAD} as
\begin{align}
    T\bar{\dot\Sigma} = \bar{J} (\mu_\nu - \mu_{\nu'}) \, .
\label{eq:EPphotochemistryI}
\end{align}
This result demonstrates that the adiabatic mechanism converts photons with high chemical potential into photons with a lower one.
Equation~\eqref{eq:EPphotochemistryI} shows that when the photons have the same chemical potential the system is detailed balanced, \textit{i.e.} the unique steady state of the dynamics is an equilibrium state, where all currents in the system vanish according to condition ~\eqref{eq:equilibrium_dyn}.
When combined with Eqs.~\eqref{eq:DissADss} and~\eqref{eq:EP_photoADfluxforce}, Eq.~\eqref{eq:EPphotochemistryI} yields the following relation
\begin{equation}
    \frac{ k_\mathrm{a}^\mathrm{E} n_\nu k_{+\Delta^\ast} \left(k_\mathrm{e}^\mathrm{Z} + k_\mathrm{se}^\mathrm{Z} n_{\nu'}\right) k_{-\Delta}}{k_\mathrm{a}^\mathrm{Z} n_{\nu'} k_{-\Delta^\ast} \left(k_\mathrm{e}^\mathrm{E} + k_\mathrm{se}^\mathrm{E} n_\nu \right)k_{+\Delta}} = \exp\left\{\frac{\mu_\nu - \mu_{\nu'}}{RT}\right\} \, ,
    \label{eq:photo-DB}
\end{equation}
which is a consequence of the local detailed balance property discussed in \S~\ref{sec:ldb}.
Importantly, this relation can be seen as a generalization of the concept known as \emph{F{\"o}ster cycle}~\cite{grabowski1977,Balzani2014}.
Indeed, by using Eqs.~\eqref{eq:munu} and~\eqref{eq:ldb_photo}, we can further recast Eq.~\eqref{eq:photo-DB} as
\begin{equation}
   \ln \frac{k_{+\Delta^\ast}}{k_{-\Delta^\ast}} - \ln \frac{k_{+\Delta}}{k_{-\Delta}} = \frac{u_\nu - u_{\nu'}}{RT} + \frac{s^\circ_\mathrm{Z^\ast} - s^\circ_\mathrm{E^\ast}}{R} - \frac{s^\circ_\mathrm{Z} - s^\circ_\mathrm{E}}{R} \, .
   \label{eq:foestercycle}
\end{equation}
F{\"o}ster's relation is recovered when assuming that $s^\circ_\mathrm{Z^\ast} - s^\circ_\mathrm{E^\ast} = s^\circ_\mathrm{Z} - s^\circ_\mathrm{E}$~\cite{foerster1950,grabowski1977,Balzani2014}, \textit{i.e.} the excess of entropy of one molecule over the other is conserved between the two electronic states.
It allows to determine the equilibrium distribution of the excited state (${k_{+\Delta^\ast}}/{k_{-\Delta^\ast}} = [\mathrm{Z^\ast}]_\mathrm{eq} / [\mathrm{E^\ast}]_\mathrm{eq}$) from the equilibrium distribution of the ground state and the excitation frequencies $\nu$ and $\nu'$, see discussion in \S~\ref{sec:discussion} for further remarks.

We conclude the discussion of the adiabatic mechanism by considering the case where photons $\gamma_\nu$ and $\gamma_{\nu'}$ comes from two different thermal sources with temperatures $T_\mathrm{r}$ and $T_\mathrm{r'}$, respectively.
In such a case, by virtue of the expression~\eqref{eq:mu_thermal} for the photons chemical potentials, the steady state entropy production in equation~\eqref{eq:EPphotochemistryI} takes the following form
\begin{align}
    \bar{\dot\Sigma} = \bar{J} \left[ u_\nu \left(\frac{1}{T} - \frac{1}{T_{\nu}}\right)  - u_{\nu'} \left(\frac{1}{T} - \frac{1}{T_{\nu'}}\right) \right]  \, .
\label{eq:EPphotochemistryIth}
\end{align}
This illustrates that the system may reach equilibrium ($\bar{\dot\Sigma} = 0$) even in the presence of a temperature difference between the two light sources ($T_\mathrm{r} \neq T_\mathrm{r'}$).
For this to happen, it suffices that the ratio of photons energies ($u_\nu / u_{\nu'}$) makes the term in square brackets vanish.
We note that detailed-balanced systems with temperature gradients are generally possible in tightly-coupled systems, where a single steady-state current (here $\bar{J}$, see Eq.~\eqref{eq:DissADss}) is responsible for the whole dissipation~\cite{esposito2009}.
Back to Eq.~\eqref{eq:EPphotochemistryIth}, when the two temperatures are equal ($T_\mathrm{r} = T_\mathrm{r'}$) ---~which happens when all the photons come from the same thermal source~--- the entropy production further reduces to
\begin{equation}
\bar{\dot\Sigma} =  \bar{J} \NA (\hbar\omega_\nu - \hbar\omega_{\nu'}) \left(\frac{1}{T} - \frac{1}{T_\mathrm{r}} \right) \, .
\label{eq:EP_thermal_AD}
\end{equation}
This shows that a nonequilibrium steady state may arise only if and only if $T \neq T_\mathrm{r}$ and the two radiative transitions couple photons with different frequencies, \textit{i.e.} $u_{\nu} \neq u_{\nu'}$.
If either of these conditions is not met, the system remains detailed balanced.

\subsection{Diabatic mechanism}
\label{sec:diabatic}

\begin{figure}[tb]
    \includegraphics[width=.50\textwidth]{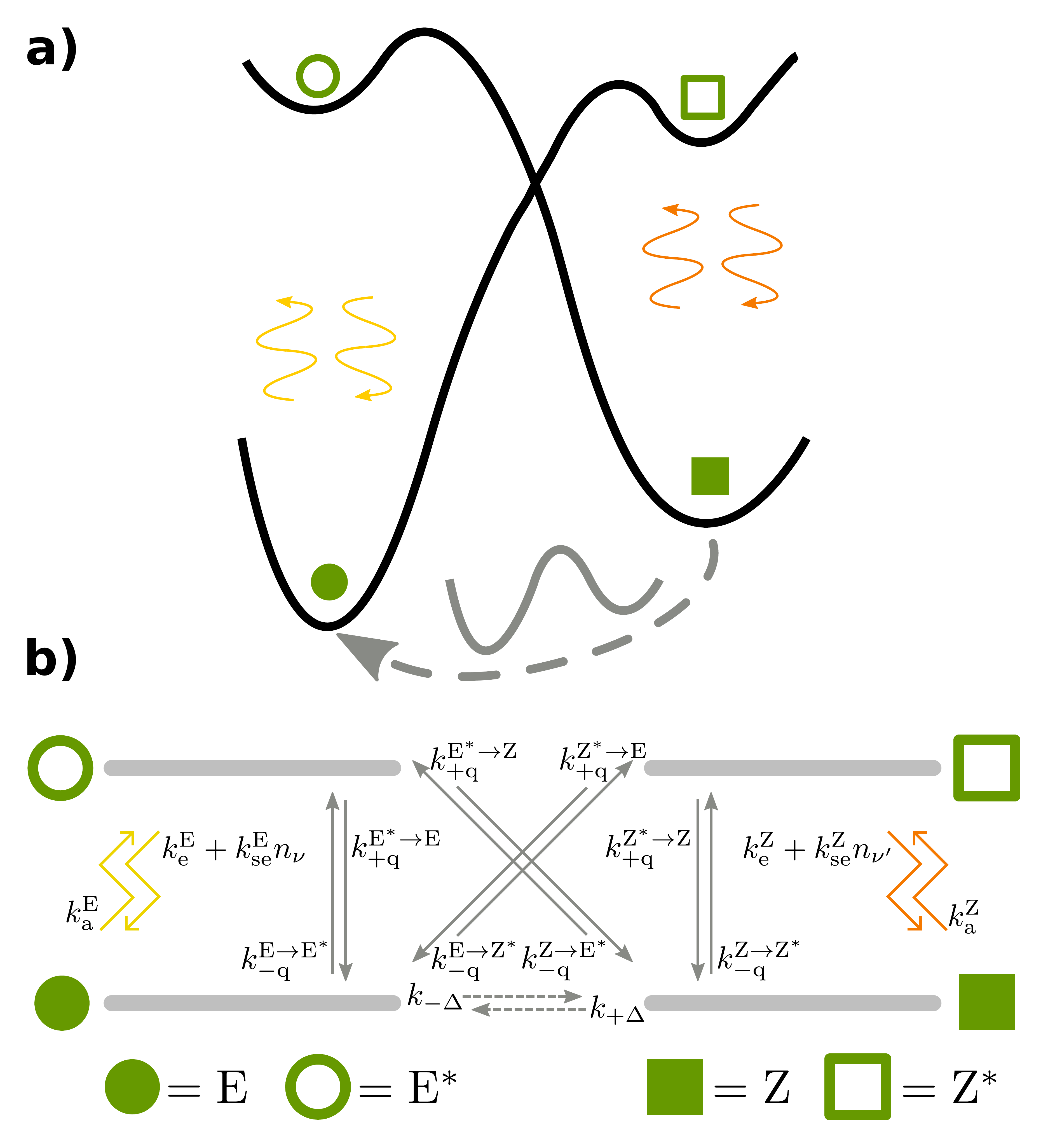}
    \centering
    \caption{\textbf{| Diabatic mechanism.}
    \textbf{a}) Pictorial representation of the ground and excited state potential energy surfaces along a reaction coordinate interconverting species $\ce{E}$ and $\ce{Z}$ through a conical intersection~\cite{Boggiopasqua2015,Persico2018}. The grey dashed arrow represents an alternative ground state pathway along a different reaction coordinate as in ref~\cite{lehn2006}.
    \textbf{b}) Schematic representation (also known as Jablonski diagram~\cite{Balzani2014}) of the mechanism.
    This scheme can represent typical E-Z photo-induced isomerizations~\cite{iupac1994} of organic molecules.~\cite{Boggiopasqua2015,Persico2018}
}
\label{FIG_photo_ci} 
\end{figure}

We now consider the photoisomerization scheme in Figure~\ref{FIG_photo_ci}.
As before, the excited states of E and Z are reached through elementary photophysical reactions, see Eqs.~\eqref{eq:adiabaticPP} and \eqref{eq:photocurrents}.
But in contrast to the adiabatic mechanism, the diabatic one includes a so-called \emph{conical intersection}~\cite{Boggiopasqua2015,Persico2018}, \textit{i.e.} the ground and excited potential energy surfaces cross each other, as in Figure \ref{FIG_photo_ci}a.
Hence, multiple quenching thermal transitions connect excited and ground state species,
\begin{align}
    \mathrm{E}^{\ast} &\ce{<=>[\mathrm{q}]} \mathrm{E} \, , \quad & \mathrm{Z}^{\ast} &\ce{<=>[\mathrm{q}]} \mathrm{Z} \\
    \mathrm{E}^{\ast} &\ce{<=>[\mathrm{q}]} \mathrm{Z} \, , \quad \text{and } & \mathrm{Z}^{\ast} &\ce{<=>[\mathrm{q}]} \mathrm{E} \, ,
    \label{eq:diabaticPP}
\end{align}
whose currents read
\begin{subequations}
    \begin{align}
     J^\mathrm{E^\ast\rightarrow \mathrm{E}}_\mathrm{q} &= k_\mathrm{+q}^\mathrm{E^\ast\rightarrow \mathrm{E}} [\mathrm{E^\ast}] - k_\mathrm{-q}^{\mathrm{E}\rightarrow \mathrm{E^\ast}} [\mathrm{E}] \\
        J^\mathrm{Z^\ast\rightarrow \mathrm{Z}}_\mathrm{q} &= k_\mathrm{+q}^\mathrm{Z^\ast\rightarrow \mathrm{Z}} [Z^\ast] - k_\mathrm{-q}^{\mathrm{Z}\rightarrow \mathrm{Z^\ast}} [\mathrm{Z}] \\
        J^\mathrm{E^\ast\rightarrow \mathrm{Z}}_\mathrm{q} &= k_\mathrm{+q}^\mathrm{E^\ast\rightarrow \mathrm{Z}} [\mathrm{E^\ast}] - k_\mathrm{-q}^{\mathrm{Z}\rightarrow \mathrm{E^\ast}} [\mathrm{Z}] \\
        J^\mathrm{Z^\ast\rightarrow \mathrm{E}}_\mathrm{q} &= k_\mathrm{+q}^\mathrm{Z^\ast\rightarrow \mathrm{E}} [\mathrm{Z}^\ast] - k_\mathrm{-q}^{\mathrm{E}\rightarrow \mathrm{Z^\ast}} [\mathrm{E}] \, .
    \end{align}
\label{eq:quenchingI}
\end{subequations}
On top of these, another non-radiative thermal reaction (represented by the dashed line in scheme~\ref{FIG_photo_ci}) converts $\mathrm{Z}$ into $\mathrm{E}$.
This conversion happens along a reaction coordinate other than the photochemical one, and its current reads
\begin{equation}
    J_\Delta = k_{+\Delta} [\mathrm{Z}] - k_{-\Delta} [\mathrm{E}] \, .
    \label{eq:termalCI}
\end{equation}
Note that, because of the local detailed balance property for thermal reactions (Eq.~\eqref{eq:ldb_thermal}), the kinetic constants appearing in Eqs.~\eqref{eq:quenchingI} and~\eqref{eq:termalCI} are not all independent from each other, and must satisfy
\begin{equation}
  \frac{k_\mathrm{-q}^{\mathrm{E}\rightarrow \mathrm{E^\ast}} k_\mathrm{+q}^\mathrm{E^\ast\rightarrow \mathrm{Z}} k_{+\Delta}}{k_\mathrm{+q}^\mathrm{E^\ast\rightarrow \mathrm{E}} k_\mathrm{-q}^{\mathrm{Z}\rightarrow \mathrm{E^\ast}} k_{-\Delta}} = \frac{ k_\mathrm{-q}^{\mathrm{Z}\rightarrow \mathrm{Z^\ast}} k_\mathrm{+q}^\mathrm{Z^\ast\rightarrow \mathrm{E}} k_{-\Delta}}{k_\mathrm{+q}^\mathrm{Z^\ast\rightarrow \mathrm{Z}} k_\mathrm{-q}^{\mathrm{E}\rightarrow \mathrm{Z^\ast}} k_{+\Delta}} = 1 \, .
\end{equation}
If such relations were violated, then nonphysical cyclic currents would originate, thus preventing the reactions from reaching detailed balance in closed systems.

The entropy production rate of the diabatic mechanism can be formally written as in Eq.~\eqref{eq:EP_photochemistryAD}.
However, in contrast to the adiabatic mechanism, the steady-state currents now must satisfy
\begin{subequations}
\begin{align}
    \bar{J}_\nu =& \bar{J}^\mathrm{E^\ast\rightarrow \mathrm{E}}_\mathrm{q} + \bar{J}^\mathrm{E^\ast\rightarrow \mathrm{Z}}_\mathrm{q} \\
    \bar{J}_{\nu'} =& \bar{J}^\mathrm{Z^\ast\rightarrow \mathrm{Z}}_\mathrm{q} + \bar{J}^\mathrm{Z^\ast\rightarrow \mathrm{E}}_\mathrm{q} \\
    \bar{J}_\Delta =& \bar{J}^\mathrm{E^\ast\rightarrow \mathrm{Z}}_\mathrm{q} - \bar{J}^\mathrm{Z^\ast\rightarrow \mathrm{E}}_\mathrm{q} = \bar{J}^\mathrm{E^\ast\rightarrow \mathrm{Z}}_\mathrm{q} - \bar{J}^\mathrm{Z^\ast\rightarrow \mathrm{E}}_\mathrm{q}\, . \label{eq:CIssIII}
\end{align}
\label{eq:CI_ss}
\end{subequations}
In particular, the two currents $\bar{J}_\nu$ and $-\bar{J}_{\nu'}$ are not constrained to be equal.
As a result, nonequilibrium steady states may arise also when the photons have the same chemical potential, giving rise to a steady-state entropy production of the form
\begin{align}
    T\bar{\dot\Sigma} = \dot{F}_\nu + \dot{F}_{\nu'} = \bar{J}_\nu \mu_\nu + \bar{J}_{\nu'} \mu_{\nu'} \, .
    \label{eq:EPphotochemistryCIss}
\end{align}

As we did for the adiabatic case (\S~\ref{sec:adiabatic}), we conclude this discussion by considering the case of photons $\gamma_\nu$ and $\gamma_{\nu'}$ coming from two different thermal sources---with temperatures $T_\mathrm{r}$ and $T_\mathrm{r'}$, respectively.
In such a case, by virtue of expression~\eqref{eq:mu_thermal} for the chemical potentials of the photons, the steady-state entropy production \eqref{eq:EPphotochemistryCIss} takes the following form
\begin{align}
    \bar{\dot\Sigma} =  \bar{J}_\nu u_\nu \left(\frac{1}{T} - \frac{1}{T_{\nu}}\right)  - \bar{J}_{\nu'} u_{\nu'} \left(\frac{1}{T} - \frac{1}{T_{\nu'}}\right)  \, .
\label{eq:EPphotochemistryCIssII}
\end{align}
As the diabatic mechanism is not tightly coupled ---~note two distinct current--force contribution at steady state~---, $\bar{\dot\Sigma}$ never vanishes if the radiation temperatures are different ($T_\mathrm{r} \neq T_\mathrm{r'}$).
When these two temperatures are equal ($T_\mathrm{r} = T_\mathrm{r'}$), or when all the photons come from the same thermal source with temperature $T_\mathrm{r}$, then the entropy production further reduces to
\begin{equation}
    \bar{\dot\Sigma} = \left(\bar{J}_\nu \NA \hbar\omega_\nu + \bar{J}_{\nu'} \NA \hbar\omega_{\nu'} \right) \left( 1 - \frac{T}{T_\mathrm{r}} \right) \, , 
    \label{eq:EPthermalphotochemistryCI}
\end{equation}
This relation shows that a photochemical process following the diabatic mechanism is detailed balanced (\textit{i.e.}, relaxes to equilibrium) only when all interacting photons have null chemical potential, namely they are in thermal equilibrium with the solution at temperature $T$ (see \S~\ref{sec:ldb}).

\subsection{Coarse-grained description}
\label{sec:cg}

To make contact with common experimental characterizations of photochemical mechanisms~\cite{Montalti2006}, one should not rely on the detailed knowledge of the excited states dynamics~\cite{Balzani2014}.
Hence, we now derive effective reactions for the photochemical mechanisms which only involve the species in the ground state, $\ce{E}$ and $\ce{Z}$.
This is achieved using a thermodynamically-consistent coarse-graining procedure specifically developed for chemical reactions networks~\cite{wachtel2018,avanzini2020cg} and based on a standard steady-state treatment of fast processes~\cite{Laidler1987}.
The underlying assumption is that the species in the excited state are sufficiently reactive for their concentrations to be considered small and always in a steady state, \textit{i.e.}
\begin{subequations}
\begin{align}
    &\dt [\mathrm{E^\ast}] =  0 \\
    &\dt [Z^\ast] =  0 \, .
\end{align}
\label{eq:CGconditions}
\end{subequations}
This time-scale-separation approximation is well justified as the life time of an excited intermediate is typically of $10^{-8}$ seconds in the absence of phosphorescent states (which are not taken into account in schemes~\ref{FIG_photochemistry_pes} and~\ref{FIG_photo_ci})~\cite{foerster1973,Balzani2014}.
Imposing Eqs.~\eqref{eq:CGconditions}, implies that the concentrations of the excited species ($[\mathrm{E^\ast}]$ and $[\mathrm{Z^\ast}]$) become functions of the concentrations of the ground state species ($[\mathrm{E}]$ and $[\mathrm{Z}]$), which thus remain the only dynamical variables in the model~\cite{Laidler1987}.
Therefore, photochemical processes are described as effective reactions involving ground state species only, and whose kinetics is given by combinations of the kinetic constants of the elementary reactions.
Standard experiments access precisely these effective kinetic parameters~\cite{Montalti2006} and we will show that   
their thermodynamic properties will depend on the underlying elementary models.

\subsubsection*{Coarse-grained adiabatic mechanism}

In the adiabatic mechanism depicted in Figure~\ref{FIG_photochemistry_pes}, conditions~\eqref{eq:CGconditions} allow the coarse-graining of the photochemical pathway $\mathrm{E} \ce{<=>} \mathrm{E^\ast} \ce{<=>} \mathrm{Z^\ast} \ce{<=>} \mathrm{Z}$ into an effective reaction of the kind $\mathrm{E} \ce{<=>} \mathrm{Z}$, whose current reads
\begin{align}
    \Psi & = k_{+\Delta^\ast} [\mathrm{E^\ast}] - k_{-\Delta^\ast} [Z^\ast] \notag \\
    & = \Phi^{\mathrm{E} \rightarrow \mathrm{Z}} k_\mathrm{a}^\mathrm{E} n_\nu [\mathrm{E}] - \Phi^{\mathrm{Z} \rightarrow \mathrm{E}} k_\mathrm{a}^\mathrm{Z} n_{\nu'} [\mathrm{Z}] \, ,
    \label{eq:pippo}
\end{align}
where
\begin{widetext}
\begin{subequations}
\begin{align}
    \Phi^{\mathrm{E} \rightarrow \mathrm{Z}} = \frac{ k_{+\Delta^\ast} (k_\mathrm{e}^\mathrm{Z}  + k_\mathrm{se}^\mathrm{Z} n_{\nu'})}{k_{-\Delta^\ast} (k_\mathrm{e}^\mathrm{E} + k_\mathrm{se}^\mathrm{E}n_\nu) + (k_{+\Delta^\ast} + k_\mathrm{e}^\mathrm{E} + k_\mathrm{se}^\mathrm{E} n_\nu) (k_\mathrm{e}^\mathrm{Z} + k_\mathrm{se}^\mathrm{Z} n_{\nu'})} \\
    \Phi^{\mathrm{Z} \rightarrow \mathrm{E}} = \frac{ k_{-\Delta^\ast}(k_\mathrm{e}^\mathrm{E} + k_\mathrm{se}^\mathrm{E} n_\nu)}{k_{-\Delta^\ast} (k_\mathrm{e}^\mathrm{E} + k_\mathrm{se}^\mathrm{E}n_\nu) + (k_{+\Delta^\ast} + k_\mathrm{e}^\mathrm{E} + k_\mathrm{se}^\mathrm{E} n_\nu) (k_\mathrm{e}^\mathrm{Z} + k_\mathrm{se}^\mathrm{Z} n_{\nu'})} \, .
\end{align}
\label{eq:phiAD}
\end{subequations}
\end{widetext}
Together with the current of the non-radiative thermal reaction \eqref{eq:ADthermocurrent}, the current \eqref{eq:pippo} fully determines the coarse-grained dynamics.
The constants $\Phi$ in Eqs.~\eqref{eq:phiAD} express the probability that, once a photon is absorbed, the photoisomerisation happens.
Importantly, these constants coincide with the ``quantum yields'' measured in experimental photochemistry, as discussed in Appendix~\ref{app:QY}.

Note that Eq.~\eqref{eq:photo-DB} can be rewritten as
\begin{align}
\frac{k_\mathrm{a}^\mathrm{E} n_\nu \, \Phi^{\mathrm{E} \rightarrow \mathrm{Z}} k_{-\Delta}}{k_\mathrm{a}^\mathrm{Z} n_{\nu'} \, \Phi^{\mathrm{Z} \rightarrow \mathrm{E}} k_{+\Delta}} = \exp\left\{\frac{\mu_\nu - \mu_{\nu'}}{RT}\right\} \, .
\label{eq:CGphoto-DB}
\end{align}
Hence, also at the coarse-grained level, we correctly predict that the system reaches equilibrium (\textit{i.e.} where all currents in the system vanish according to condition ~\eqref{eq:equilibrium_dyn}) when the photons have the same chemical potential.
When this is not the case, the system reaches a nonequilibrium steady state where
\begin{equation}
    \bar{J} = \bar{\Psi} = \bar{J}_\Delta \, .
    \label{eq:ADCGss}
\end{equation}
As predicted for tightly coupled reaction mechanisms \cite{wachtel2018,avanzini2020cg}, the single effective reaction obeys the local detailed balance, as one can see from Eq.~\eqref{eq:CGphoto-DB}.
Indeed, the coarse-grained photochemical affinity reads
\begin{equation}
\hspace{-2em}
    \mathcal{A} \equiv \mu_\mathrm{E} + \mu_\nu - \mu_\mathrm{Z} - \mu_{\nu'} =
    RT\ln \left\{\frac{k_\mathrm{a}^\mathrm{E} n_\nu \, \Phi^{\mathrm{E} \rightarrow \mathrm{Z}}  [\mathrm{E}]}{k_\mathrm{a}^\mathrm{Z} n_{\nu'} \, \Phi^{\mathrm{Z} \rightarrow \mathrm{E}} [\mathrm{Z}]}\right\} \, ,
    \label{eq:ldb_ADCG}
\end{equation}
and the coarse-grained entropy production can be written as
\begin{align}
    \dot\Sigma^\mathrm{cg} =& \Psi \frac{\mathcal{A}}{T}  + J_{\Delta} \frac{A_\Delta}{T} \, .
    \label{eq:photochemistry_CG_EP}
\end{align}
Provided that the time scale separation hypothesis leading to condition~\eqref{eq:CGconditions} holds, Eq.~\eqref{eq:photochemistry_CG_EP} coincides with Eq.~\eqref{eq:EP_photoADfluxforce}, and the full entropy production can be quantified from the coarse-grained description: $\dot\Sigma^\mathrm{cg} = \dot\Sigma$~\cite{avanzini2020cg}.
At steady state, where Eq.~\eqref{eq:ADCGss} applies, we find
\begin{align}
    T \bar{\dot\Sigma}^\mathrm{cg} = T \bar{\dot\Sigma} = \bar{J} (\mu_\nu - \mu_{\nu'}) \, ,
    \label{eq:CG_EPss}
\end{align}
which exactly coincides with Eq.~\eqref{eq:EPphotochemistryI} regardless of how good is the time scale separation hypothesis~\cite{wachtel2018}.

\subsubsection*{Coarse-grained diabatic mechanism}

In the diabatic mechanism depicted in Figure~\ref{FIG_photo_ci}, conditions~\eqref{eq:CGconditions} allow the coarse-graining of the four photochemical pathways $\mathrm{E} \ce{<=>} \mathrm{E^\ast} \ce{<=>} \mathrm{Z}$, $\mathrm{E} \ce{<=>} \mathrm{E^\ast} \ce{<=>} \mathrm{E}$, $\mathrm{Z} \ce{<=>} \mathrm{Z^\ast} \ce{<=>} \mathrm{E}$, and $\mathrm{Z} \ce{<=>} \mathrm{Z^\ast} \ce{<=>} \mathrm{Z}$ into four effective reactions of the kind $\mathrm{E} \ce{<=>} \mathrm{Z}$, $\mathrm{E} \ce{<=>} \mathrm{E}$, $\mathrm{Z} \ce{<=>} \mathrm{E}$, and $\mathrm{Z} \ce{<=>} \mathrm{Z}$, respectively.
The corresponding coarse-grained currents read
\begin{subequations}
\begin{align}
    \Psi^{\mathrm{E}\rightarrow \mathrm{Z}}_\nu  &= \Phi^\mathrm{E^\ast\rightarrow \mathrm{Z}} k_\mathrm{a}^\mathrm{E} n_\nu [\mathrm{E}] - \Gamma^{\mathrm{E}\rightarrow \mathrm{Z}}_\nu  \\
    \Psi^{\mathrm{E}\rightarrow \mathrm{E}}_\nu  &= \Phi^\mathrm{E^\ast\rightarrow \mathrm{E}} k_\mathrm{a}^\mathrm{E} n_\nu [\mathrm{E}] - \Gamma^{\mathrm{E}\rightarrow \mathrm{E}}_\nu  \\
    \Psi^{\mathrm{Z}\rightarrow \mathrm{E}}_{\nu'}  &= \Phi^\mathrm{Z^\ast\rightarrow \mathrm{E}} k_\mathrm{a}^\mathrm{Z} n_{\nu'} [\mathrm{Z}] - \Gamma^{\mathrm{Z}\rightarrow \mathrm{E}}_{\nu'}  \\
    \Psi^{\mathrm{Z}\rightarrow \mathrm{Z}}_{\nu'}  &= \Phi^\mathrm{Z^\ast\rightarrow \mathrm{Z}} k_\mathrm{a}^\mathrm{E} n_{\nu'} [\mathrm{Z}] - \Gamma^{\mathrm{Z}\rightarrow \mathrm{Z}}_{\nu'} \, .
\end{align}
\label{eq:CGcurrentsCI}
\end{subequations}
All these four effective reactions coarse-grain pathways in which either $\mathrm{E}$ or $\mathrm{Z}$ absorb a photon and then thermally quench towards either the same or the isomeric ground state.
As before, the $\Phi$ terms express the quantum yields of the various processes, while the $\Gamma$ terms are proportional to the (very small) backward-quenching thermal rates and are therefore experimentally negligible.
Expressions of both $\Phi$s and $\Gamma$s  are reported in Appendix~\ref{app:CGCI}.
The overall net rate of $\mathrm{E}$ to $\mathrm{Z}$ conversion due to photochemical processes is given by the difference
\begin{align}
\hspace{-1em}
    \Psi^{\mathrm{E}\rightarrow \mathrm{Z}}_\nu - \Psi^{\mathrm{Z}\rightarrow \mathrm{E}}_{\nu'} \approx \Phi^\mathrm{E^\ast\rightarrow \mathrm{Z}} k_\mathrm{a}^\mathrm{E} n_\nu [\mathrm{E}] - \Phi^\mathrm{Z^\ast\rightarrow \mathrm{E}} k_\mathrm{a}^\mathrm{Z} n_{\nu'} [\mathrm{Z}] \, ,
    \label{eq:CGphotocurrentA_ci}
\end{align}
while the net absorption currents can be expressed as
\begin{subequations}
\begin{align}
    J_{\nu} &= \Psi^{\mathrm{E}\rightarrow \mathrm{Z}}_\nu +  \Psi^{\mathrm{E}\rightarrow \mathrm{E}}_\nu \approx (\Phi^\mathrm{E^\ast\rightarrow \mathrm{Z}} + \Phi^\mathrm{E^\ast\rightarrow \mathrm{E}}) k_\mathrm{a}^\mathrm{E} n_\nu [\mathrm{E}] \\
    J_{\nu'} &= \Psi^{\mathrm{Z}\rightarrow \mathrm{E}}_{\nu'} +  \Psi^{\mathrm{Z}\rightarrow \mathrm{Z}}_{\nu'} \approx (\Phi^\mathrm{Z^\ast\rightarrow \mathrm{E}} + \Phi^\mathrm{Z^\ast\rightarrow \mathrm{Z}}) k_\mathrm{a}^\mathrm{Z} n_{\nu'} [\mathrm{Z}] \, .
\end{align}
\label{eq:CGabscurrentCI}
\end{subequations}
At steady state, the following equality holds
\begin{equation}
    \bar{J}_\Delta =  \bar{\Psi}^{\mathrm{E}\rightarrow \mathrm{Z}}_\nu - \bar{\Psi}^{\mathrm{Z}\rightarrow \mathrm{E}}_{\nu'} \, .
    \label{eq:CICGss}
\end{equation}

In contrast with tightly coupled mechanisms, coarse-grained reactions~\eqref{eq:CGcurrentsCI} for the diabatic mechanism do not satisfy any local detailed balance property.
Thus, their thermodynamic affinities can be solely defined in terms of chemical potentials, and they do not equate the log ratio of forward and backward currents, \textit{cf.} Eq.~\eqref{eq:ldb_ADCG}:
\begin{align}
    \mathcal{A}^{\mathrm{E}\rightarrow \mathrm{Z}}_\nu =& \mu_\mathrm{E} + \mu_\nu - \mu_\mathrm{Z} \, , \quad \mathcal{A}^{\mathrm{E}\rightarrow \mathrm{E}}_\nu = \mu_\nu \nonumber \\
    \mathcal{A}^{\mathrm{Z}\rightarrow \mathrm{E}}_{\nu'} =& \mu_\mathrm{Z} + \mu_{\nu'} - \mu_\mathrm{E} \, , \quad \mathcal{A}^{\mathrm{Z}\rightarrow \mathrm{Z}}_{\nu'} = \mu_{\nu'} \, .
\end{align}
In analogy with Eq.~\eqref{eq:photochemistry_CG_EP}, the entropy production at the coarse-grained level reads~\cite{wachtel2018,avanzini2020cg}
\begin{multline}
    \dot{\Sigma}^\mathrm{cg} = J_\Delta \frac{A_\Delta}{T} + \Psi^{\mathrm{E}\rightarrow \mathrm{Z}}_\nu \frac{\mathcal{A}^{\mathrm{E}\rightarrow \mathrm{Z}}_\nu}{T}  + \Psi^{\mathrm{Z}\rightarrow \mathrm{E}}_{\nu'} \frac{\mathcal{A}^{\mathrm{Z}\rightarrow \mathrm{E}}_{\nu'}}{T} \\
    + \Psi^{\mathrm{E}\rightarrow \mathrm{E}}_\nu \frac{\mathcal{A}^{\mathrm{E}\rightarrow \mathrm{E}}_\nu}{T} + \Psi^{\mathrm{Z}\rightarrow \mathrm{Z}}_{\nu'} \frac{\mathcal{A}^{\mathrm{Z}\rightarrow \mathrm{Z}}_{\nu'}}{T} \, .
    \label{eq:EPphotochemistryCI_CG}
\end{multline}
At the steady state, using Eqs.~\eqref{eq:CGabscurrentCI} and \eqref{eq:CICGss}, we find
\begin{align}
    T\bar{\dot\Sigma}^\mathrm{cg} & = \left(\bar{\Psi}^{\mathrm{E}\rightarrow \mathrm{Z}}_\nu +  \bar{\Psi}^{\mathrm{E}\rightarrow \mathrm{E}}_\nu\right) \mu_\nu + \left(\bar{\Psi}^{\mathrm{Z}\rightarrow \mathrm{E}}_{\nu'} +  \bar{\Psi}^{\mathrm{Z}\rightarrow \mathrm{Z}}_{\nu'}\right) \mu_{\nu'} \nonumber \\
    & = \bar{J}_\nu \mu_\nu + \bar{J}_{\nu'} \mu_{\nu'} \, .
    \label{eq:EPphotochemistryCI_CGss}
\end{align}
As for the adiabatic mechanism, the steady-state coarse-grained entropy production (Eq.~\eqref{eq:EPphotochemistryCI_CGss}) coincides with the fine-grained one (Eq.~\eqref{eq:EPphotochemistryCIss})~\cite{wachtel2018}, while in transient regimes the effectiveness of the coarse-grained estimation depends on the validity of the time scale separation approximation, Eq.~\eqref{eq:CGconditions}~\cite{avanzini2020cg}.

\section{Conclusions and Discussion}
\label{sec:discussion}

In this paper, we introduced a modern formulation of nonequilibrium thermodynamics for systems made of chemicals interacting with incoherent radiation.
Our description leverages the advances made for purely-thermal chemical-reactions systems in isothermal ideal dilute solutions \cite{polettini2014,rao2016}.

In Sec.~\ref{sec:absEmProc}, we focused on a simple photophysical mechanism in which photons may excite a chemical species.
By using the property of local detailed balance, \textit{viz.} Einstein relations, we established a rigorous and thermodynamically-consistent description of such excitation dynamics.
The main outcome of our description follows as we show that the nonequilibrium free energy of chemicals plus radiation $F$, Eq.~\eqref{eq:noneqF}, acts as a Lyapunov function for a closed system relaxing to thermal equilibrium, \S \ref{sec:closedsystem}:
$F$ never increases and reaches its minimum at thermal equilibrium.
Mathematically, this proves that the equilibrium state is globally stable.

While we focused on chemical systems, the approach developed in \S~\ref{sec:thermopot} is general and can be used to study the thermodynamics of any kind of light-reacting matter, provided that the local equilibrium assumption holds.
This means that the sole mechanism generating dissipation are the reacting events mediating changes in the molar concentration of photons and molecules. In absence of reacting events the molecules are at equilibrium in solution and the photons are at (possibly another) equilibrium in the volume occupied by the solution. 
Our treatment is also consistent with previous results obtained in more specific contexts, including semiconductor physics~\cite{Wurfel2016,Milazzo1997}.

The results derived in Sec.~\ref{sec:photochemistry} on basic photochemical mechanisms describing light-induced conversion of a chemical species ($\mathrm{E}$) into another one ($\mathrm{Z}$) are compatible with both experimental observations and previous theoretical descriptions.
The adiabatic mechanism analyzed in \S~\ref{sec:adiabatic} is usually employed to model excited-state proton-transfer reactions, where $\mathrm{E}$, $\mathrm{Z}$ and their excited counterparts differ by the location of an acidic proton~\cite{grabowski1977,formosinho1993,Balzani2014}.
As first pointed out by F{\"o}ster~\cite{foerster1950}, a shift in the acid-base equilibrium distribution between $\mathrm{E}$ and $\mathrm{Z}$ under light irradiation from a single source cannot be reached if the two excitation frequencies $\nu$ and $\nu'$ are equal~\cite{grabowski1977,formosinho1993, Balzani2014}. 
This observation is predicted by Eq.~\eqref{eq:EPphotochemistryI} in the general case and by Eq.~\eqref{eq:EP_thermal_AD} when the light source is a thermal one.
Indeed, in both cases there's no steady-state entropy production when $\nu = \nu'$, meaning that the system will stay at the acid-base equilibrium when irradiated.
Also, a generalization of the well-known concept of F{\"o}ster cycle~\cite{foerster1950,grabowski1977} is offered by Eq.~\eqref{eq:photo-DB}, which allows to link ratios between the kinetic constants of the thermally-induced non-radiative steps with the differences in excitation frequencies and standard entropies of the chemical species involved.

The diabatic mechanism analyzed in \S~\ref{sec:diabatic} models the most common mechanism for photochemical E-Z isomerizations, where $\mathrm{E}$ and $\mathrm{Z}$ are for instance two stable configurations of an alkene or an azobenzene derivative~\cite{Balzani2014,Boggiopasqua2015,Persico2018}.
A major biological example of this kind of processes is the primary event in vision \cite{cerullo2010}.
For an example drawn from synthetic chemistry, we refer to the minimalist two-steps light-driven molecular motor proposed in Ref.~\cite{lehn2006}, whose first experimental implementation is based on imine groups \cite{lehn2014,lehn2015} and for which the scheme in Fig.~\ref{FIG_photo_ci} is a realistic model.
Contrary to the adiabatic case, we demonstrated that a photoisomerization following the diabatic mechanism can be driven out of equilibrium with only one frequency of light.
This is in line with common models describing synthetic light-driven molecular motors~\cite{feringa2009,sabatino2019}.

The typically very high reactivity of excited intermediates in fine-grained photochemical reaction schemes makes the coarse-grained level of description developed in \S~\ref{sec:cg} the most natural framework to compare theory with experiments.
Indeed, coarse-grained expressions~\eqref{eq:pippo} and~\eqref{eq:CGphotocurrentA_ci} are usually employed to fit experimental data and measure quantum yields~\cite{Montalti2006}.
Crucially, despite the fact that the coarse-grained currents~\eqref{eq:pippo} and~\eqref{eq:CGphotocurrentA_ci} read identically, only for the adiabatic mechanism one can estimate its dissipative contribution ---~the affinity~---- using the local detailed balance property at the level of coarse-grained fluxes, see Eq.~\eqref{eq:ldb_ADCG}.
Contrary, for photochemical processes following the diabatic mechanism (the most common mechanism for photo-isomerizations~\cite{Balzani2014,Boggiopasqua2015}), the sole knowledge of the quantum yields appearing in the coarse-grained net current~\eqref{eq:CGphotocurrentA_ci} is not enough to estimate the dissipation.
Instead, one needs to measure also the net absorption currents in Eq.~\eqref{eq:CGabscurrentCI}, which allow to apply either Eq.~\eqref{eq:EPphotochemistryCI_CG} or Eq.~\eqref{eq:EPphotochemistryCI_CGss} for computing the dissipation.

\section{Perspectives}
\label{sec:conclusions}

Our results constitute a step forward towards a unified nonequilibrium thermodynamics for chemical systems powered by different free-energy sources such as chemostats and light. They also open the way to performance studies on free-energy transduction and storage in finite-time and far from equilibrium which were inaccessible before.

Building on the results developed in this paper, we plan next to generalize our thermodynamic theory to continuous absorption/emission spectra, as well as elementary bimolecular photophysical (\textit{e.g.} bimolecular quenchings) and photochemical (\textit{e.g.} photodissociations) light-induced processes.
Another limitation to overcome in order to model real experiments is to go beyond the assumption that the radiostats are able to fix the same molar concentration of photons in the whole system.
While this is a good approximation for ideal dilute solutions irradiated uniformly, photochemical experiments are usually conducted with light coming from one side of the system.
This may give rise to nonlinear shielding effect~\cite{micheau1992}.
We anticipate that such extensions will allow to study interesting nonequilibrium phenomena such as bistability and oscillations in photochemical systems~\cite{nitzan1973,micheau1992,gentili2021}, whose thermodynamic cost and constraints have never been compared with the cost of chemically driven oscillations~\cite{avanzini2019cw}.

\begin{acknowledgments}
    The authors are thankful to F. Avanzini and D. Accomasso for insightful discussions.
    EP and ME acknowledge funding from the European Research Council, project NanoThermo (ERC-2015-CoG Agreement No.681456).
    RR was funded by \textit{The Martin A. and Helen Chooljian} Membership in Biology at the Institute for Advanced Study.
\end{acknowledgments}


\appendix
\section{On the quantum yield}
\label{app:QY}
The standard textbook definition~\cite{Balzani2014,Montalti2006} of quantum yield for the direct $\mathrm{E^\ast} \rightarrow \mathrm{Z^\ast} \rightarrow \mathrm{Z}$ process reads:
\begin{align}
    \Phi^{\mathrm{E} \rightarrow \mathrm{Z}}_0 = \frac{k_{+}^\ast}{k_{+}^\ast + (k_\mathrm{e}^\mathrm{E} + k_\mathrm{se}^\mathrm{E} n_\nu)} \frac{(k_\mathrm{e}^\mathrm{Z} + k_\mathrm{se}^\mathrm{Z} n_{\nu'})}{k_{-}^\ast + (k_\mathrm{e}^\mathrm{Z} + k_\mathrm{se}^\mathrm{Z} n_{\nu'})} \, ,
\end{align}
but we can also consider an indirect process of the kind $\mathrm{E^\ast} \rightarrow \mathrm{Z^\ast} \rightarrow \mathrm{E^\ast} \rightarrow \mathrm{Z^\ast} \rightarrow \mathrm{Z}$, for which the quantum yield reads
\begin{align}
    \Phi^{\mathrm{E} \rightarrow \mathrm{Z}}_1 = \Phi^{\mathrm{E} \rightarrow \mathrm{Z}}_0 \frac{k_{-}^\ast}{k_{-}^\ast + (k_\mathrm{e}^\mathrm{Z} + k_\mathrm{se}^\mathrm{Z} n_{\nu'})} \frac{k_{+}^\ast}{k_{+}^\ast + (k_\mathrm{e}^\mathrm{E} + k_\mathrm{se}^\mathrm{E} n_\nu)} \, .
\end{align}
The global quantum yield of the photoisomerisation is therefore the sum of the quantum yields of all the possible paths going from $\mathrm{E^\ast}$ to $\mathrm{Z}$.
By taking all of them into account, we correctly find:
\begin{align}
    &\Phi^{\mathrm{E} \rightarrow \mathrm{Z}} = \sum_{i=0}^\infty \Phi^{\mathrm{E} \rightarrow \mathrm{Z}}_i = \\ \nonumber
    &= \Phi^{\mathrm{E} \rightarrow \mathrm{Z}}_0 \sum_{i=0}^\infty \left( \frac{k_{-}^\ast}{k_{-}^\ast + (k_\mathrm{e}^\mathrm{Z} + k_\mathrm{se}^\mathrm{Z} n_{\nu'})} \frac{k_{+}^\ast}{k_{+}^\ast + (k_\mathrm{e}^\mathrm{E} + k_\mathrm{se}^\mathrm{E} n_\nu)}\right)^i 
\end{align}
\vspace{0mm}

\section{Coarse-grained $\Phi$'s and $\Gamma$'s terms for the diabatic mechanism}
\label{app:CGCI}
The quantum yields in the coarse-graned version of the diabatic mechanism read
\begin{subequations}
\begin{align}
    \Phi^\mathrm{E^\ast\rightarrow \mathrm{Z}} &= \frac{k_\mathrm{+q}^\mathrm{E^\ast\rightarrow \mathrm{Z}}}{k_\mathrm{+q}^\mathrm{E^\ast\rightarrow \mathrm{Z}} + k_\mathrm{+q}^\mathrm{E^\ast\rightarrow \mathrm{E}} + (k_\mathrm{e}^\mathrm{E} + k_\mathrm{se}^\mathrm{E} n_\nu)} \, ,\\
    \Phi^\mathrm{E^\ast\rightarrow \mathrm{E}} &= \frac{k_\mathrm{+q}^\mathrm{E^\ast\rightarrow \mathrm{E}}}{k_\mathrm{+q}^\mathrm{E^\ast\rightarrow \mathrm{Z}} + k_\mathrm{+q}^\mathrm{E^\ast\rightarrow \mathrm{E}} + (k_\mathrm{e}^\mathrm{E} + k_\mathrm{se}^\mathrm{E} n_\nu)} \, ,\\
    \Phi^\mathrm{Z^\ast\rightarrow \mathrm{E}} &= \frac{k_\mathrm{+q}^\mathrm{Z^\ast\rightarrow \mathrm{E}}}{k_\mathrm{+q}^\mathrm{Z^\ast\rightarrow \mathrm{E}} + k_\mathrm{+q}^\mathrm{Z^\ast\rightarrow \mathrm{Z}} + (k_\mathrm{e}^\mathrm{Z} + k_\mathrm{se}^\mathrm{Z} n_{\nu'})} \, ,\\
    \Phi^\mathrm{Z^\ast\rightarrow \mathrm{Z}} &= \frac{k_\mathrm{+q}^\mathrm{Z^\ast\rightarrow \mathrm{Z}}}{k_\mathrm{+q}^\mathrm{Z^\ast\rightarrow \mathrm{E}} + k_\mathrm{+q}^\mathrm{Z^\ast\rightarrow \mathrm{Z}} + (k_\mathrm{e}^\mathrm{Z} + k_\mathrm{se}^\mathrm{Z} n_{\nu'})} \, ,
\end{align}
\end{subequations}
where $k_\mathrm{q}$'s are the quenching thermal rates going from one exited species to its ground state isomer.
Quantum yields here express the probabilities that, given that either $\mathrm{E}$ or $\mathrm{Z}$ absorbed a photon, the photoisomerisation happens or the excited state quenches back to the initial ground state.
The $\Gamma$ terms are proportional to the (very small) backward quenching thermal rates:
\begin{subequations}
\begin{align}
    \Gamma^{\mathrm{E}\rightarrow \mathrm{Z}}_\nu &= \frac{[\mathrm{Z}]k_\mathrm{-q}^{\mathrm{Z}\rightarrow \mathrm{E^\ast}} (k_\mathrm{e}^\mathrm{E} + k_\mathrm{se}^\mathrm{E} n_\nu + k_\mathrm{+q}^\mathrm{E^\ast\rightarrow \mathrm{E}}) - [\mathrm{E}]k_\mathrm{-q}^{\mathrm{E}\rightarrow \mathrm{E^\ast}} k_\mathrm{+q}^\mathrm{E^\ast\rightarrow \mathrm{Z}}}{k_\mathrm{+q}^\mathrm{E^\ast\rightarrow \mathrm{Z}} + k_\mathrm{+q}^\mathrm{E^\ast\rightarrow \mathrm{E}} + (k_\mathrm{e}^\mathrm{E} + k_\mathrm{se}^\mathrm{E} n_\nu)} \, , \\
    \Gamma^{\mathrm{E}\rightarrow \mathrm{E}}_\nu &= \frac{[\mathrm{E}]k_\mathrm{-q}^{\mathrm{E}\rightarrow \mathrm{E^\ast}} (k_\mathrm{e}^\mathrm{E} + k_\mathrm{se}^\mathrm{E} n_\nu + k_\mathrm{+q}^\mathrm{E^\ast\rightarrow \mathrm{Z}}) - [\mathrm{Z}]k_\mathrm{-q}^{\mathrm{Z}\rightarrow \mathrm{E^\ast}}k_\mathrm{+q}^\mathrm{E^\ast\rightarrow \mathrm{E}}}{k_\mathrm{+q}^\mathrm{E^\ast\rightarrow \mathrm{Z}} + k_\mathrm{+q}^\mathrm{E^\ast\rightarrow \mathrm{E}} + (k_\mathrm{e}^\mathrm{E} + k_\mathrm{se}^\mathrm{E} n_\nu)} \, , \\
    \Gamma^{\mathrm{Z}\rightarrow \mathrm{E}}_\nu &= \frac{[\mathrm{E}]k_\mathrm{-q}^{\mathrm{E}\rightarrow \mathrm{Z^\ast}} (k_\mathrm{e}^\mathrm{Z} + k_\mathrm{se}^\mathrm{Z} n_\nu + k_\mathrm{+q}^\mathrm{Z^\ast\rightarrow \mathrm{Z}}) - [\mathrm{Z}]k_\mathrm{-q}^{\mathrm{Z}\rightarrow \mathrm{Z^\ast}} k_\mathrm{+q}^\mathrm{Z^\ast\rightarrow \mathrm{E}}}{k_\mathrm{+q}^\mathrm{Z^\ast\rightarrow \mathrm{E}} + k_\mathrm{+q}^\mathrm{Z^\ast\rightarrow \mathrm{Z}} + (k_\mathrm{e}^\mathrm{Z} + k_\mathrm{se}^\mathrm{Z} n_\nu)} \, , \\
    \Gamma^{\mathrm{Z}\rightarrow \mathrm{Z}}_\nu &= \frac{[\mathrm{Z}]k_\mathrm{-q}^{\mathrm{Z}\rightarrow \mathrm{Z^\ast}} (k_\mathrm{e}^\mathrm{Z} + k_\mathrm{se}^\mathrm{Z} n_\nu + k_\mathrm{+q}^\mathrm{Z^\ast\rightarrow \mathrm{E}}) - [\mathrm{E}]k_\mathrm{-q}^{\mathrm{E}\rightarrow \mathrm{Z^\ast}}k_\mathrm{+q}^\mathrm{Z^\ast\rightarrow \mathrm{Z}}}{k_\mathrm{+q}^\mathrm{Z^\ast\rightarrow \mathrm{E}} + k_\mathrm{+q}^\mathrm{Z^\ast\rightarrow \mathrm{Z}} + (k_\mathrm{e}^\mathrm{Z} + k_\mathrm{se}^\mathrm{Z} n_\nu)} \, ,
\end{align}
\end{subequations}
and are therefore not experimentally accessible with standard techniques.


%
\end{document}